\documentclass[]{aa}
\usepackage{xcolor}

\usepackage[varg]{txfonts}

\usepackage{graphicx}
\graphicspath{{../plots/}}
\DeclareGraphicsExtensions{.pdf,.jpeg,.png,.jpg,.eps}

\usepackage{wasysym}

\usepackage{twoopt}
\usepackage{hyperref}
\hypersetup{pdfauthor=Christoph Mordasini, pdftitle=Planetary Evolution with Atmospheric Photoevaporation II, breaklinks=true, colorlinks=true, urlcolor=blue, linkcolor=blue, citecolor=blue, bookmarksopen=true}

\bibpunct{(}{)}{;}{a}{}{,} 

\defcitealias{2020A&A...638A..52M}{Paper I}

\makeatletter
\renewcommand*\aa@pageof{, page~\thepage{} of~\pageref*{LastPage}}
\makeatother

\title{
    Planetary evolution with atmospheric photoevaporation II\@:\\
    Fitting the slope of the radius valley by combining boil-off and XUV-driven escape
}

\author{
    L. Affolter \inst{\ref{inst:unibe}}\thanks{\emph{Current address:} Institute for Particle Physics and Astrophysics, ETH Zurich, 8093 Zurich, Switzerland} \and
    C. Mordasini \inst{\ref{inst:unibe}} \and
    A.~V. Oza \inst{\ref{inst:jpl},\ref{inst:unibe}} \and
    D. Kubyshkina \inst{\ref{inst:aasgraz},\ref{inst:trinity}} \and
    L. Fossati \inst{\ref{inst:aasgraz}}
}

\institute{
    Physics Institute, University of Bern, Sidlerstrasse 5, CH-3012 Bern, Switzerland\label{inst:unibe}\and
    Jet Propulsion Laboratory, California Institute of Technology, Pasadena, California\label{inst:jpl}\and
    Space Research Institute, Austrian Academy of Sciences, Schmiedlstrasse 6, A-8042 Graz, Austria\label{inst:aasgraz}\and
    School of Physics, Trinity College Dublin, the University of Dublin, College Green, Dublin-2, Ireland\label{inst:trinity}
}

\date{Received 13.09.2021 / Accepted 28.06.2023}
\abstract{
Observations by the Kepler satellite have revealed a  {gap} between larger sub-Neptunes and smaller super-Earths that atmospheric escape models had predicted as an evaporation valley prior to the discovery.
}
{
{We seek to contrast results from a simple XUV-driven energy-limited  escape model against  those from a direct hydrodynamic  model.} The latter calculates the thermospheric temperature structure self-consistently, including cooling effects like thermal conduction. Besides XUV-driven escape, it also includes the boil-off escape regime where the escape is driven by the atmospheric thermal energy and low planetary gravity, catalysed by stellar continuum irradiation. We couple these two escape models to an internal structure model and follow the planets' temporal evolution.
}
{
To examine the population-wide imprint of the two escape models and to compare to observations, we {first} employ a  {rectangular} grid, tracking the evolution of planets  {as a function of core mass and orbital period} over gigayear timescales.  {We then study the slope of the valley {also} for initial conditions derived from the observed Kepler planet population.}
}
{
 {For the rectangular grid,} we find that the power-law slope of the valley with respect to orbital period is \(-0.18\) and \(-0.11\) in the energy-limited and hydrodynamic model, respectively.  {For the initial conditions derived from the  Kepler planets, the results are similar (\(-0.16\) and \(-0.10\)).}
While the  {slope found with the energy-limited model} is steeper than observed, the  {one of the hydrodynamic model} is in excellent agreement with observations.
The reason for the shallower slope is caused by the two regimes in which the energy-limited approximation fails. First, low-mass planets at low-to-intermediate stellar irradiation. For them, boil-off dominates mass loss. However, boil-off is absent in the energy-limited model, thus it underestimates escape relative to the hydrodynamic model. Second, massive  compact planets at high  {XUV irradiation}. For them, the energy-limited approximation overestimates escape relative to the hydrodynamic model because of cooling by thermal conduction, which is neglected in the {energy-limited model}.
}
{
The two effects act together in concert to yield in the hydrodynamic model a shallower slope of the valley that agrees very well with observations. We conclude that a hydrodynamic escape model that includes boil-off and a more realistic treatment of cooling mechanisms can reproduce one of the most important constraints for escape models, the valley slope.
}


\titlerunning{Planet evolution with atmospheric escape II\@: fitting the valley slope}
\authorrunning{Affolter et al.}

\begin{document}

\maketitle
\section{Introduction}
The analysis of Kepler satellite data has revealed a dearth of \(1.9\,R_\oplus \) planets, often referred to as a valley or gap between the two populations of sub-Neptunes and super-Earths \citep{2017AJ....154..109F,2018MNRAS.479.4786V}.
Atmospheric escape models had predicted this dearth as an {evaporation valley} prior to the observational discovery \citep{2013ApJ...776....2L,2013ApJ...775..105O,2014ApJ...795...65J}.
While the properties of the valley are now observationally quite well known, its origin is still debated.
The leading hypotheses are XUV-driven atmospheric photoevaporation \citep{2013ApJ...775..105O,2014ApJ...795...65J} and atmospheric loss driven by the cooling of the core \citep{2019MNRAS.487...24G}.
Alternatively, it might also be a direct imprint of formation, separating dry planets that have formed inside the ice line from volatile-rich ones that have migrated in from beyond the ice line \citep{2020A&A...643L...1V,izidoro2022},  {or} a consequence of impact-driven atmospheric erosion \citep{10.1093/mnras/stz3052}. {It might also} be caused by primordial gas accretion alone \citep{leekaralis2022}.

But even within the context of just the atmospheric escape models, the details of the atmospheric mass loss driving the evolution, like the different escape regimes (boil-off, blow-off, and {J}eans escape) and limiting physical processes like energy- or radiation-recombination-limited {escape} are still poorly understood \citep[for recent reviews on atmospheric escape, see][]{2008SSRv..139..355J,2017ApJ...843..122Z,2019AREPS..47...67O}.

This complexity is not surprising as escaping atmospheres in the Solar System \citep[e.g.,][]{2010Icar..210....1L} are also not yet fully understood in spite of in-situ observations due to the complexity of molecular kinetic interactions which include hydrodynamically escaping atmospheres.
It is important to note that the main escape mechanisms in the Solar System are of non-thermal nature while for close-in exoplanets, thermal escape mechanisms dominate.

Examples of atmospheric escape in the Solar System include hydrodynamic escape in the Earth's H exosphere \citep{1975JGR....80.2247B}, on early Earth and Venus \citep{1981Icar...48..150W}, and on Titan \citep{2016Icar..271..202J} and a blend of Jeans and energy-limited escape on the Kuiper Belt Objects \citep{2015ApJ...809...43J} as well as plasma-driven escape on Mars \citep{2018Icar..315..146J,2019GeoRL..46.4144L} and Mercury \citep{2018P&SS..159...97G,2004jpsm.book..561T}.

While the latter is also conceivable at a close-in irradiated exoplanet  {system \citep{2019ApJ...885..168O,2020MNRAS.497.5271G}}, especially given the correlation in X-ray luminosity \citep{2019ApJ...876...22M}, here we focus on the former mechanism, thermally driven hydrodynamic escape. We note that there is a growing interest in assessing also the atmospheric escape of young bodies like protoplanets where the atmosphere is thought to be sourced directly from a magma ocean \citep[e.g.][]{2021Icar..36414451C}.

Energy-limited escape has long been used to approximate hydrodynamic escape to first order \citep{1981Icar...48..150W}.
It has the advantage of simplicity, hiding the complex physics in the evaporation efficiency factor \(\eta_\text{XUV}\).
Especially in planet evolution calculations, \(\eta_\text{XUV}\) is often assumed to be a constant, in contrast to the results of direct hydrodynamic simulations \citep[e.g.,][]{2016A&A...585L...2S,2017ApJ...847...29O}.
Nevertheless, the energy-limited approximation can fare rather well compared to some extensive kinetic simulations \citep{2013ApJ...768L...4J} up to a critical threshold in the  {reduced heating rate}. Above this threshold, when the escape transitions to a more Jeans-like regime \citep{2011ApJ...729L..24V}, the energy-limited escape approximation overestimates the escape rate by orders of magnitude \citep{2015A&A...576A..42S}.
A second limitation is that at high  {EUV fluxes}, the escape becomes radiation-recombination-limited rather than energy-limited \citep{2009ApJ...693...23M}.
Third, even at intermediate  {XUV fluxes}, the energy-limited approximation is not applicable for planets with a particularly high or low gravitational potential \citep{2021A&A...650A..94K}.
Finally, in the initial evolutionary phase of planets immediately after the dissipation of the natal protoplanetary gas disc, escape of primordial H/He envelopes is driven by a combination of low gravity and high atmospheric temperatures.
This leads to very vigorous boil-off \citep{2015A&A...576A..87S,2016ApJ...817..107O,2017A&A...598A..90F}, which is also neglected in a purely XUV-driven, energy-limited approach.

Therefore, in light of recent observations characterising the radius valley  {\citep{2018MNRAS.479.4786V,2019ApJ...875...29M,petigura2022,dichang2022,hovaneylen2023}}, we seek to statistically test both the energy-/recombination-limited model and a direct numerical treatment of hydrodynamic escape \citep{2018A&A...619A.151K} {which} overcomes the assumptions and limitations of the energy-limited formula, against these observational constraints.
We thus here work under the assumption that the valley is a consequence of atmospheric escape.

Regarding the observational data, we will make use of the analysis of the California--Kepler Survey in tandem with parallaxes from the Gaia mission \citep{2018A&A...616A...1G} by \citet{2019ApJ...875...29M}  {and \citet{petigura2022}}.
The latest Gaia release is essential, as the new parallaxes provide a more accurate determination of planetary radii on a population-wide scale.
We also compare to the observations from \citet{2018MNRAS.479.4786V}  {which} predates the second GAIA data release, but uses astroseismology to determine accurate stellar parameters.  {Finally, we also use the valley locus as determined by \citet{hovaneylen2023} based on short  cadence Kepler data.}

Our paper is organised as follows: in Sect.~\ref{sec:model} we describe our theoretical model and describe the simulation setup in Sect.~\ref{sect:procedure}.
In Sect.~\ref{sec:results},  {using a rectangular grid of initial conditions}, we compare the locus and slope of the evaporation valley in a radius {--}period diagram and demonstrate on a population-wide level how the hydrodynamic escape model, but not the energy-/recombination-limited  {model} lead to excellent agreement with the observ{ed} slope. The  {physical reason for this} will become clear in  {Sects.}~\ref{subect:evoindividualcases} and~\ref{sect:compevapmodelsongrid} where we study selected individual evolutionary tracks which highlight the differences between the two models, and compare the models on the entire grid, respectively. 

 {In Sect.~\ref{sect:keplercomp}, we determine the slope of the valley as found with the two evaporation models using now initial conditions derived from the observed Kepler population instead of the rectangular grid.} We end our paper with a summary and the conclusions (Sect.~\ref{sec:discussion}).  {In Appendix~\ref{appendix:lumi} we address the impact of the post-formation entropy on the valley locus.}

\section{Model}\label{sec:model}
Our approach to modelling planetary evolution under the effect of atmospheric escape is two-fold: on the one hand, we use a simpler semi-analytical energy and radiation-recombination limited escape model of XUV-driven atmospheric photoevaporation.
This escape model was already used in the first paper of the series, \citet{2020A&A...638A..52M}.
Here, we slightly update it, as described later in this section.
The model is itself based on \citet{2014ApJ...795...65J,2018ApJ...853..163J}.
On the other hand, we now also use the tabulated escape rates obtained with a sophisticated numerical hydrodynamic escape model \citep{2018A&A...619A.151K}.
In both cases, we couple these escape models to our model of temporal planetary interior evolution (cooling and contraction).

We then use this model to evolve a population of close-in planets. 
In both approaches, the interior evolution component of the calculations are performed with the \texttt{COMPLETO21} planet evolution model that was described in details in the first paper \citep[][hereafter Paper I]{2020A&A...638A..52M}.

This evolution model simulates the temporal thermodynamical and compositional evolution of the planet by solving the classical 1D spherically symmetric interior structure equations.
The planets consist of an iron/silicate core described with the EOS of \citet{2007ApJ...669.1279S} and a H/He envelope described with the EOS of \citet{1995ApJS...99..713S}.
The atmosphere is described with an improved version of the double grey model of \citet{2010A&A...520A..27G}, as described in \citet{2014ApJ...795...65J}.
This interior structure model yields, together with the mass loss, in particular the radius of the planet as a function of time as well as the remaining H/He envelope mass.
In this work, as stated, we implement a new coupling to the hydrodynamic escape model described in \citet{2018A&A...619A.151K} that we shall next summarise along with the standard energy-/recombination-limited model that we used previously.

\subsection{Energy-/recombination-limited escape model}\label{sec:elim}
Energy-limited (EL) escape assumes that the energy is lost most efficiently by gas expansion to space rather than conduction (downwards) or radiation (upwards to space).
An in-depth analysis of the assumptions underlying the EL formalism can be found in \citet{2021A&A...650A..94K}, while a detailed description of our energy-/recombination-limited escape model is given in \citetalias{2020A&A...638A..52M}.

In its simplest form, EL escape is written as
\begin{equation}
    \dot{M}_\mathrm{EL} \mathbin{\sim} \frac{Q(R_\mathrm{a})}{U(R_\mathrm{p})},
\end{equation}
where \(U = G M_\mathrm{p} /R_\mathrm{p}\) is the specific binding energy of matter in the potential well of a planet of mass \(M_\mathrm{p}\) and radius \(R_\mathrm{p}\) with \(G\) the gravitational constant. \(R_\mathrm{a}\) is the effective radius at which incoming radiation is absorbed on the planet.
In our model, the radius where the EUV radiation is absorbed is calculated as described in \citet{2009ApJ...693...23M}.

The complexity then arises in the heating rate \(Q\), which is assumed to occur in the upper atmosphere.
For XUV-driven escape, we can thus approximate the energy-limited escape due to upper atmospheric heating as
\begin{equation}
    \label{eq:elim}
    \dot{M}_\mathrm{U} = \eta_\text{XUV} \frac{\pi R_\mathrm{p} R_\mathrm{a}^2 F_\text{XUV}}{G M_\mathrm{p} K_\text{tide}},
\end{equation}
where we assume the only energy absorbed by the planetary envelope cross-section \(\pi R_\mathrm{a}^2\) driving escape is stellar X-ray and EUV radiation (collectively: XUV) which is written as the flux at the planet's position \(F_\text{XUV}\).
It is further assumed that only a fraction of the total flux of the star drives mass loss, given by the evaporative efficiency factor \(\eta_\text{XUV}\).
This efficiency factor is, as discussed above, problematic as it oversimplifies the heating and cooling specific to each planet.
Finally, the \(K_\text{tide}\) factor corrects for the gravity due to the stellar tide described in \citet{2007A&A...472..329E}.

In contrast to \citetalias{2020A&A...638A..52M}, where a constant \(\eta_\text{XUV}\) was assumed, we now use an \(\eta_\text{XUV}\) that depends on the escape speed \(v_\mathrm{esc}\),  {as suggested by approximate fits to the mass-loss simulations of \citet{2012MNRAS.425.2931O}. The same functional form was also used in \citet{ 2017ApJ...847...29O,rogersowen2021}}.
 {For the specific values of the parameters in Eq.~\ref{eq:eta}, we use the ones which were found in \citet{2019ApJ...874...91W} to lead to the best reproduction of the observed Kepler planet population.} 
\begin{equation}
    \label{eq:eta}
     \eta_\text{XUV} = 0.17 \, {\left( \frac{v_\mathrm{esc}}{23\,\mathrm{km\,s\textsuperscript{\mbox{-1}}}} \right)}^{-0.42}
\end{equation}
 {These values are consistent with the ones found  in \citet{rogersowen2021}.}

In practice, it is  {however} found that \(\eta_\text{XUV}\) remains in the range of 0.1--0.3 because of the small exponent, and because the escape speed does not change by orders of magnitudes.
Consequently, the differences to a simulation with a constant \(\eta_\text{XUV}\) are {very limited}. {In particular, the slope of the valley is not affected by this modification and remains virtually identical to the value found in \citetalias{2020A&A...638A..52M} with a constant \(\eta_\text{XUV}\). In this paper, it was also found that fixed globally higher or lower values of \(\eta_\text{XUV}\) do not affect the slope, but rather shift the valley up and down as a whole. This behavior is in turn in perfect agreement with the predictions of the analytical model derived in \citetalias{2020A&A...638A..52M} (see Eq. 36 in that work).}

As described in \citet{2014ApJ...795...65J} and \citet{2018ApJ...853..163J}, heating by UV and X-rays are treated separately in the model, using the criterion of \citet{2012MNRAS.425.2931O} to identify the dominant process.

In the radiation recombination-limited (RR) regime \citep{2009ApJ...693...23M} that occurs at high  {EUV fluxes}, the escape rate is given by the equilibrium of photoionisation with radiative recombination.
In this regime, we closely follow \citet{2009ApJ...693...23M} to calculate the escape rate.
The final escape rate is taken to be the minimum of the energy-limited and the recombination-limited escape rates \citep{2017MNRAS.472..245L}.

The fact that the numerical results obtained with this evaporation model can be very well understood with an analytical model based on the energy-limited formula only \citepalias{2020A&A...638A..52M}, shows that the importance of the recombination-limited regime is small for the planets studied here.

\subsection{Hydrodynamic escape model}\label{sec:hydro}
To estimate atmospheric escape within a more sophisticated direct hydrodynamic approach, we employ the grid of planetary upper atmosphere models presented by \citet{2018A&A...619A.151K}.
The grid consists of roughly 7000 models, each corresponding to a planet, and covers the following parameter space: planetary mass (\(M_\mathrm{p}\)) between 1 and 39 Earth masses; planetary radius (\(R_\mathrm{p}\)) between 1 and 10 Earth radii; planetary equilibrium temperature (\(T_\mathrm{eq}\)) between \(300\,\mathrm{K}\) and \(2000\,\mathrm{K}\);\@ stellar mass between \(0.4\) and \(1.3\) solar masses; and XUV flux between \(0.4\) and \(10^4\) the one experienced by present Earth because of solar irradiation, with values scaled for the specific stellar masses.
The range of orbital separations covered by the grid was set on the basis of the stellar mass and planetary equilibrium temperature, thus stellar radius (\(R_*\)) and effective temperature (\(T_\text{eff}\)).
\(R_*\) and \(T_\text{eff}\) were derived considering the range of radii and effective temperatures covered by a star of each considered mass along the main-sequence on the basis of stellar evolutionary tracks \citep{2001ApJS..136..417Y}.
Considering all stellar masses, the orbital separation ranges between \(0.002\) and \(1.3\,\mathrm{AU}\).

The basic hydrodynamic model used to construct the grid is an updated version of the model developed in \citet{2016MNRAS.460.1300E}.
It considers a pure hydrogen atmosphere subject to heating and cooling processes{, 
including radiative Ly-\(\alpha \) cooling following \citet{yelle2004} and \(\text{H}_3^+\) cooling following \citet{miller2013} as well as adiabatic cooling \citep[see details in ][]{2018A&A...619A.151K}.} {These cooling processes are not explicitly included in the energy-limited approximation but are {(incorrectly) supposed to be captured}  in the {(constant)} efficiency factor \(\eta_\text{XUV}\) introduced in Sect.~\ref{sec:elim}.}

The model  {numerically} solves a full set of hydrodynamic equations, including  {energy and momentum} conservation laws and continuity equations accounting for the full atmospheric hydrogen chemistry comprising dissociation, recombination, and ionisation. The complete list of chemical reactions is given in \citet{2018A&A...619A.151K}.
The model does not account for the presence of metals, which could induce additional heating and/or cooling that have been shown to be effective for ultra-hot primary \citep{fossati2021}, secondary \citep[e.g. Earth-like;][]{2018A&A...617A.107J} and rock vapour \citep{2021MNRAS.502..750I} atmospheres. However, conditions at planets in the grid are such that {condensation may occur} in the lower atmosphere, limiting the penetration of heavy elements in the upper atmosphere \citep[see, e.g.,][]{2021A&A...646A.171C}. {At the early evolution stages, when the extreme atmospheric escape takes place, the small amount of metals in a hydrogen-dominated upper atmosphere will be dragged away by the hydrogen outflow and will have little impact on the atmospheric mass-loss rates \citep{1986Icar...68..462Z,1987Icar...69..532H,2018Icar..307..327O,2020SSRv..216...74L}, while at the later stages, metal abundances can be fractionated by more moderate hydrodynamic escape or non-thermal escape processes \citep[e.g., ][]{2020JGRA..12527639G}. Even more importantly, one has also to consider that metal abundances may vary significantly between individual planets, already from the formation stage. Therefore, our current knowledge remains limited regarding the precise composition of sub-Neptunes, and the uncertainty in metal abundances does not enable one to place reasonable assumptions valid for planets spanning over a wide parameter space. Furthermore, the possibility to include metals into consideration is limited from the practical numerical side; the computational costs of the hydrodynamic models allowing for a proper metal treatment (i.e., including a detailed chemical framework and a photoionisation treatment including the explicit calculation of the energy levels populations) are at the moment still too high for computing a large and dense grid of mass loss rates similar to that used in the present study.}

The boundaries of the model are the photospheric radius of the planet (lower boundary) and its Roche lobe (upper boundary)%
\begin{equation}
    \label{eq:roche}
    R_\text{roche} = a {\left[ \frac{M_\mathrm{p}}{3 (M_\mathrm{p} + M_*)} \right]}^{1/3},
\end{equation}%
where \(a\) is the planet's orbital distance and \(M_*\) the stellar mass. The model accounts for stellar heating in two wavelength intervals: EUV and X-ray ranges, assuming that the integrated flux of each range is emitted at a single wavelength (\(60\) and \(5\,\mathrm{nm}\), respectively).
The heating is included into the energy conservation equation as an external source given by%
\begin{equation}\label{eq:hydro_heating}
    Q_\mathrm{m} = \frac{1}{2} \eta^*_\text{XUV} \, \sigma_\mathrm{m} \, n_\text{H} \int^{\frac{\pi}{2}+\arccos\left(\frac{1}{r}\right)}_{0} {J_\mathrm{m}(r,\theta) \cdot \sin(\theta)} \, \mathrm{d}\theta,
\end{equation}%
where \(m\) stands for either X-ray or EUV radiation, \(\sigma_\mathrm{m}\) is an absorption cross-section of hydrogen for the specific wavelength, \(n_\text{H}\) is the hydrogen (H + H$_{2}$) density, \(r\) is the distance to the planetary centre, and \(J_\mathrm{m}(r,\theta)\) is a function with spherical coordinates describing the spacial variations of the XUV flux due to atmospheric absorption%
\begin{equation}\label{eq:hydro_vol_heat_func}
    J_\mathrm{m}(r, \theta) = \exp\left({-\int^{R_\text{roche}}_{r} \sigma_\mathrm{m} \, n_\text{H}(\xi) \sqrt{\xi^2 - r^2 \sin(\theta)} \, \xi \, \mathrm{d}\xi}\right),
\end{equation}%
which is approximately equivalent to the optical depth at \(\theta = 0\).

{We note that the \(\eta^*_\text{XUV}\) in Equation~\ref{eq:hydro_heating} {for the hydrodynamic model} is not the same as \(\eta_\text{XUV}\) given by Equation~\ref{eq:eta} {for the energy-limited model}. 
This is because \(\eta^*_\text{XUV}\) does not account for any additional cooling processes {or other physical mechanisms supposed to be captured by (or hidden in) \(\eta_\text{XUV}\),} as they are included self-consistently in the {hydrodynamic} model  \citep{2018A&A...619A.151K}. {Instead, \(\eta^*_\text{XUV}\) accounts solely for the efficiency of the photoionisation heating, and is not an overall evaporation efficiency as \(\eta_\text{XUV}\) in the energy-limited model.}

Given that a self-consistent calculation of 
{\(\eta^*_\text{XUV}\)} is currently too time-consuming for computing a large grid, it was set to be equal to a constant value of 15\%, which is a reasonable assumption for the considered \(M_\mathrm{p}\) range \citep[e.g.,][]{2014A&A...571A..94S,2016A&A...585L...2S}. {We note that despite this compelled simplification, the hydrodynamic code remains a superior model relative to the energy-limited approximation, as the latter approach relies on many more assumptions than just a constant heating efficiency and the absence of the explicitly modelled radiative cooling processes. In particular, it omits the contribution from the thermal energy of the planet atmosphere and the stellar VIS/IR irradiation (as discussed below) and makes crude assumptions on the atmospheric structure, which is calculated self-consistently by the hydrodynamic model \citep[for a more thorough discussion, see][]{2021A&A...650A..94K}.}

{Typically, for hydrodynamic planetary/stellar wind models, the initially subsonic outflow (we set the bulk velocity \(V_\text{bulk}\) equal to zero at the lower boundary) is accelerated to supersonic velocities before the flow reaches the Roche lobe. Within our grid of models, it happens typically at a distance of a few planetary radii. To ensure that the atmospheres of planets in the grid remain collisional throughout the simulation, we calculate the Knudsen number for each point of the atmospheric profiles a posteriori. The atmospheric mass-loss rate is finally defined as the flow through the sphere of radius \(r\) in a unit of time (\(m_\text{H}n_\text{H}(r)V_\text{bulk}(r)\)) multiplied by the surface of this sphere. As the outflow is continuous, for the computation of the mass-loss rate the specific distance \(r\) is not relevant (except for the small region at the lower boundary), but for convenience it is taken at the Roche radius.}

 {The predictions of our model are comparable to those made by other hydrodynamic models, including the more sophisticated ones (such as those calculating self-consistently the heating efficiency in various approaches as \citealt{2009ApJ...693...23M} and \citealt{2016A&A...585L...2S}, and models accounting for the detailed spectral energy distribution as \citealt{2016ApJ...818..107G}, or 3D geometry as \citealt{2021MNRAS.500.3382C}). Further details about the physical model and the grid, including the comparison to observations and to the results of other literature models, can be found in \citet{2018A&A...619A.151K}.}

By construction, the hydrodynamic model accounts for Jeans escape, XUV hydrodynamic escape, and boil-off escape regimes.
This is, as we shall see below, of central importance for the results for the valley found here. The model also transitions smoothly from one escape regime to the other depending on the system parameters.
 {To ease distinguishing between the latter two regimes, it is convenient to employ the restricted Jeans parameter \citep{2017A&A...598A..90F}, which is a combination of the physical planetary parameters and is defined as}
\begin{equation}
    \label{eq:lambda}
    \Lambda = \frac{G M_\mathrm{p} m_\text{H}}{k_\mathrm{B} T_\mathrm{eq} R_\mathrm{p}},
\end{equation}
with \(m_\text{H}\) the mass of the hydrogen atom and \(k_\mathrm{B}\) the Boltzmann constant.

Planets with a \(\Lambda \) smaller than 15--35 are in the boil-off regime, where the escape is driven by the atmospheric thermal energy and low planetary gravity \citep{2016ApJ...817..107O,2017A&A...598A..90F,2016ApJ...825...29G}. The specific critical value depends on the stellar mass and orbital separation.
Such planets are typically just released from the protoplanetary gas disk.
A \(\Lambda \) of 20 is for an isothermal gas identical to the condition derived by \citet{2016ApJ...817..107O} for the occurrence of boil-off, namely that the planet radius is larger than about 0.1 times the Bondi radius.

For the boil-off regime, it is crucial that the hydrodynamic model also accounts for the stellar continuum (dominated by VIS/IR) heating that can drive escape, in contrast to the energy-/radiation-limited model that is driven by  {XUV heating} only.
The continuum heating is implicitly included by fixing the temperature at the lower boundary equal to \(T_\mathrm{eq}\).
We have verified that the photospheric temperatures of our model planets as predicted by the interior structure model is always very close to \(T_\mathrm{eq}\).
The largest difference (a temperature that is about 4\% higher) occurs for the most massive planets we model at the beginning of the simulations, which is due to the contribution of the intrinsic luminosity.
Overall, the difference is, however, much smaller and generally less than 1\%.

Studies comparing the hydrodynamical model used here with the energy-limited escape have found the following \citep{2018A&A...619A.151K,2021A&A...650A..94K}: For planets with \(\Lambda \) less than about 20, the energy-limited formalism on average severely underestimates mass-loss rates, because it lacks the continuum (VIS/IR) heating.
For higher \(\Lambda \), the energy-limited rate provides an upper limit on the mass-loss rate, with significant overestimations possible depending on a planet's gravitational potential.

Model outputs for each planet in the grid are profiles of the main atmospheric parameters, which allow deriving the effective radii of the stellar XUV absorption, and atmospheric escape rates.
To finally obtain the mass-loss rates for any planet during its evolution, we linearly interpolate among the grid points.
Our interpolation scheme is simpler than the one in \citet{2018A&A...619A.151K}, but allows to fully exploit all grid data including the borders of the tabulated regions.

\subsection{Stellar XUV luminosity as a function of time}
A modification of our theoretical model relative to \citetalias{2020A&A...638A..52M} is the usage of a more recent description of the stellar XUV luminosity as a function of time.
In \citetalias{2020A&A...638A..52M}, we used the data of \citet{2005ApJ...622..680R}.
In the updated model, we use instead \citet{2019ApJ...876...22M}.
These authors compiled observationally derived relations extracted from several studies \citep{2012MNRAS.422.2024J,2014AJ....148...64S} of the X-ray luminosity of stars as a function of time and stellar type.
We use their mean X-ray luminosity as function of time \(L_\mathrm{X}(t)\) and convert it into the extreme UV-luminosity \(L_\text{EUV}(t)\) with the relation of \citet{2011A&A...532A...6S}.

\begin{figure}
	\centering
    \includegraphics[width=\linewidth]{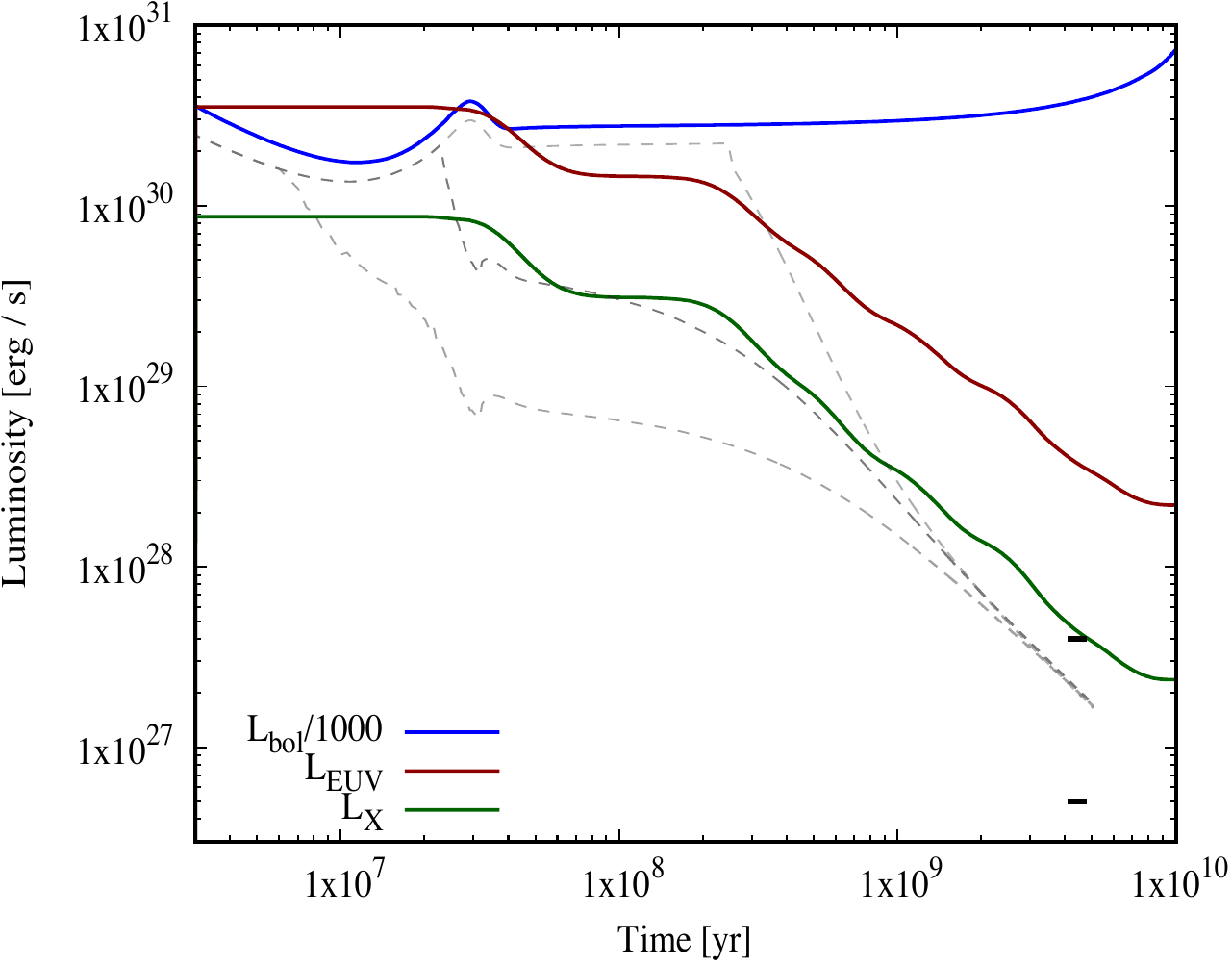}
    \caption{
        Temporal evolution of the bolometric (blue), X-ray (green), and EUV luminosity (brown) of \(1\,\mathrm{M}_\odot \) star as assumed in our model.
        The bolometric luminosity is divided by a factor 1000 to bring it on a similar scale as \(L_\mathrm{X}\) and \(L_\text{EUV}\).
        The two black bars near \(4.5\,\mathrm{Gyr}\) show the range of our Sun's \(L_\mathrm{X}\) over the course of a solar cycle.
        The grey dashed lines show for comparison the \(L_\mathrm{X}\) of \citet{2015A&A...577L...3T} for the 10th, 50th, and 90th percentiles of the rotational distribution.
    }\label{fig:lumis}
\end{figure}

Figure~\ref{fig:lumis} shows \(L_\mathrm{X}\), \(L_\text{EUV}\), and the bolometric luminosity \(L_\mathrm{bol}\) of a \(1\,M_\odot \) star in our model.
At young ages, the XUV luminosity is on the order of \(10^{-3}\) the bolometric luminosity, as expected \citep[e.g.,][]{2020SSRv..216..143G}, and approximately constant in time, except for a certain drop at around the time (\(40\,\mathrm{Myr}\)) when the star reaches the main sequence.
Afterwards, it decreases approximately following a power law.
At \(4.6\,\mathrm{Gyr}\), the predicted \(L_\mathrm{X}\) is compatible with the Sun's measured \(L_\mathrm{X}\) at its activity maximum.

We compare our model with the one of \citet{2015A&A...577L...3T}.
They calculated the \(L_\mathrm{X}\) predicted for stars on the 10th, 50th, and 90th percentiles of the stellar rotational distribution.
We see that our relation is similar to their 50th percentile case.
At the earliest epochs, our \(L_\mathrm{X}\) is about a factor 2 lower than theirs, while at high ages, the fall off is a bit faster in \citet{2015A&A...577L...3T}.
The impact of varying the stellar XUV on the locus of the evaporation valley was studied in \citetalias{2020A&A...638A..52M}  {(see also \citealt{ketzerpoppenhaeger2023})}.

\section{Procedure}\label{sect:procedure}
 {We use the model to evolve a large number of close-in low-mass planets. To set their initial (post-formation) properties, we follow two approaches: a rectangular grid, and initial conditions derived from the planetary population detected of the Kepler satellite. The rectangular grid allows us to see clearly (but also under idealised assumptions) the population-wide imprints of the two evaporation models.  The initial conditions derived from the Kepler population give us an understanding if these imprints remain visible also when the initial conditions are more complex, in particular when there is a spread in the post-formation envelope mass at a fixed core mass.}

\subsection{ {Rectangular} grid of models and initial conditions}
Our  {first approach} is the same as in \citetalias{2020A&A...638A..52M}: for the two escape models, we simulated a  {rectangular} grid of 6000 planets each, equally spaced in semi-major axis \(a\) and core mass \(M_\text{core}\) ranging from \(0.01\) to \(0.6\,\mathrm{AU}\) in \(0.01\,\mathrm{AU}\) increments and from \(1\) to \(20\,M_\oplus \) in \(0.2\,M_\oplus \) increments.
We let these planets evolve from \(3\,\mathrm{Myr}\)  {(a typical lifetime of protoplanetary disks, \citealt{mamajek2009})} to \(10\,\mathrm{Gyr}\) around a \(1\,\mathrm{M}_\odot \) star. With this data, we can analyse the slope and temporal evolution of the evaporation valley.

The initial (post-formation) H/He envelope mass \(M_\mathrm{env,0}\) was estimated as
\begin{equation}\label{eq:menve0}
M_\mathrm{env,0} = 0.024\,M_\oplus \, {\left( \frac{M_\text{core}}{1\,M_\oplus} \right)}^{2.23} {\left( \frac{a}{1\,\mathrm{AU}} \right)}^{0.72}.
\end{equation}
As described in \citetalias{2020A&A...638A..52M}, this relation was  {found as a typical mean value} from planet formation simulations by \citet{2014A&A...566A.141M} based on the core accretion paradigm which find the envelope mass similarly as in \citet{1996Icar..124...62P}, but include many additional effects like orbital migration and disk evolution.
The results of \citetalias{2020A&A...638A..52M} employing the XUV-driven energy-/recombination-limited escape model indicate, however, that using a different initial envelope mass within plausible ranges should not strongly influence the location of the valley. An additional argument for this weak dependency in the context of boil-off is that if the initial envelope mass is larger, then the radius is larger and thus the boil-off escape is larger as well. Therefore, at the end of the boil-off phase, an initially larger and an initially smaller planet end up with a similar radius because the larger planet had a stronger escape compared to the smaller one \citep{2020MNRAS.499...77K}.  {We nevertheless investigate the impact of the initial envelope mass further in Sect.~\ref{sect:keplercomp}.} The initial intrinsic luminosity of the planets was also estimated as a function of core and envelope mass based on the same formation simulations. This is the same approach as in \citetalias{2020A&A...638A..52M}.  {In Appendix~\ref{appendix:lumi} we study the impact of different post-formation luminosities/entropies, finding an only weak influence on the valley locus.}

Regarding their bulk composition, all cores have an Earth-like composition with a 2:1 silicate-to-iron mass fraction.
Such a composition is in agreement with the composition of planets below the valley \citep{2017ApJ...847...29O,2018ApJ...853..163J}.

\subsection{ {Initial conditions derived from Kepler survey}}\label{Sect:compKeplerinit}
\begin{figure*} 
	\centering
    \includegraphics[width=\linewidth]{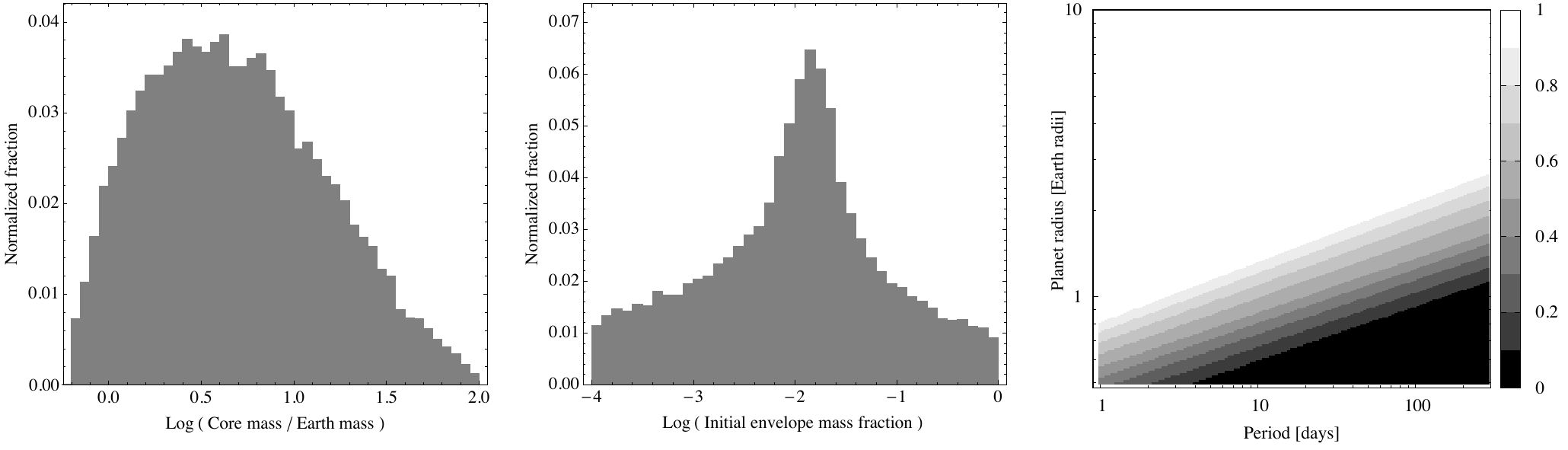}
    \caption{ {Left and middle panel: histogram of the distribution of the core masses and initial (post-formation) envelope mass fractions \(M_\mathrm{env,0}/M_\text{core}\) for the comparison with the Kepler planet population. Both these distributions are taken from 
        \citet{rogersowen2021}. The right panel shows the average detection probability of the Kepler satellite \(p_\mathrm{det}\) as function of orbital period and planet radius \citep{petiguramarcy2018}. The final probability to find a planet is given by \(p_\mathrm{det}\) times its geometric transit probability \(p_\mathrm{tr}\).} 
    }\label{fig:ics}
\end{figure*}
 {The initial conditions on the rectangular grid are not tuned to reproduce the observed Kepler planet population, which makes the comparison with observations less straightforward. The grid also assumes in an idealised way that there is a unique value of the initial envelope mass as a function of core mass and orbital distance. Formation models \citep[e.g.,][]{2014A&A...566A.141M}, but also inference analyses of the post-formation properties of the Kepler planets \citep{rogersowen2021}, indicate in contrast a spread in post-formation envelope masses.}

 {Our second approach for the initial conditions is thus to adopt distributions for the orbital period, core mass, and initial envelope mass that have been derived from fitting the observed properties of the close-in low-mass population found by the Kepler satellite through inference analyses \citep{guptaschlichting2020,rogersowen2021}. In these works, the core and initial envelope mass distributions that were derived lead --- after evolution under the effects of core-powered mass loss and photoevaporation, respectively --- to a synthetic population that agrees with the observed period-radius distribution (the CKS data, \citealt{fultonpetigura2018}). The period distribution is also derived from the Kepler observations and given as \citep{rogersgupta2021}
\begin{equation}\label{eq:perioddist}
    \frac{\mathrm{d}N}{\mathrm{d}\log P} \propto \begin{cases}
        P^{2}, & {P < 8\; \mathrm{days}}, \\
        \mathrm{constant}, & {P > 8\; \mathrm{days}}.
    \end{cases}  
\end{equation}
The core mass distribution we adopt is the one inferred in \citet{rogersowen2021} in their preferred Model III\@. It peaks at a core mass of about \(4\,M_\oplus \), with a tail extending to about \(100\,M_\oplus\). The post-formation envelope mass fraction is also taken from this source. It is a distribution peaking at an envelope mass fraction of about 4\%, but covering a significant range. In contrast to the theoretical relation (Eq.~\ref{eq:menve0}), the envelope mass fraction is here an independent quantity. Both these distributions are shown in the left and middle panel of Fig.~\ref{fig:ics}. With these initial conditions, we calculated the evolution of 37242 and 37416 planets from \(3\,\mathrm{Myr}\) to \(10\,\mathrm{Gyr}\) for the energy-limited and hydrodynamic evaporation model, respectively.}

 {To understand if the imprints of the different evaporation models remain observable, we apply a simple synthetic detection bias of the Kepler satellite to the model output. In this way, we get the detectable synthetic population. For each synthetic planet, we compute the detectability as a function of planet size and orbital period. It has two components. The first component is the geometric transit probability \(p_\mathrm{tr}\). For it, following \citet{petiguramarcy2018}, we use that a randomly inclined planet on a circular orbit transits with an impact parameter \(b < 0.9\) with a probability \(p_\mathrm{tr} = 0.9 R_\odot/a\), where \(R_\odot \) is the radius of the Sun (we only consider \(1 \, M_\odot \) stars in this paper). The second component is the detection probability \(p_\mathrm{det}\), which depends mainly on the S/N of the observations. Here we take the average \(p_\mathrm{det}\) also from \citet{petiguramarcy2018}, which is based on the transit injection and recovery study of \citet{christiansenclarke2015}. This is shown in the right panel of Fig.~\ref{fig:ics}. The total probability is then the product of the two probabilities, \(p_\mathrm{tr} \times p_\mathrm{det}\). By comparing \(p_\mathrm{tr} \times p_\mathrm{det}\) with a random number drawn from the standard uniform deviate, we obtain the detectable synthetic planets. To have enough detectable synthetic planets despite the low detection probability of the transit method, we oversample 100 times, i.e. we run through the list of synthetic planets 100 times, obtaining each time different detectable planets. This means that the same planet can end up several times in the final list of detectable planets. However, for the statistical analysis at hand, this is not an issue. In this way, we end up with 114634 and 113465 detectable synthetic planets for the energy-limited and hydrodynamic model, respectively.}

 {The initial condition distributions derived in \citet{guptaschlichting2020,rogersowen2021} lead with their relative theoretical forward models to synthetic populations agreeing with the Kepler data. However, the distributions that the two works infer differ from each other. This reflects that these `fitting' initial conditions are also a function of the forward model. Here, we use again another forward model (or even two, counting the two different evaporation models). Thus, we cannot expect that we will find with our forward model in the end a detectable subpopulation agreeing equally well with the actual Kepler population. As we will see in Sect.~\ref{sect:keplercomp}, our synthetic detectable populations do, however, still share key properties with the observed population, like for example a bimodal radius distribution. We could, in principle, conduct a similar hierarchical inference process as \citet{guptaschlichting2020,rogersowen2021} to derive our own fitting initial conditions. However, practically this would be difficult because of the much higher computational cost of our forward model compared to theirs. It would also be beyond the scope of this paper, which addresses the comparison of two evaporation models.}

\subsection{Quantifying the locus of the valley}\label{sect:quantifyingvalleylocus}
Following the approach of several previous papers  {\citep[e.g.,][]{2018MNRAS.479.4786V,2018MNRAS.479.5303L,2018ApJ...853..163J,2019ApJ...875...29M,rogersgupta2021,petigura2022,hovaneylen2023}}, we quantify the valley locus with a power law.
Normalising at an orbital period of 10 days, we express  {for the rectangular grid simulations} the planetary radius \(R_\mathrm{p}\) of the largest bare core (i.e., most massive planet which has completely lost its H/He envelope) at a given orbital period \(p\) as
\begin{equation}
    \label{eq:rbare}
    R_\mathrm{b}(p) = \tilde{R}_\mathrm{b} {\left( \frac{p}{10\,\mathrm{days}} \right)}^\alpha
\end{equation}
where \(\tilde{R}_\mathrm{b}\) is the value at 10 days and the slope is
\begin{equation}
	\alpha = \frac{\mathrm{d}\log{R_\mathrm{p}}}{\mathrm{d}\log{p}}.
\end{equation}
Using this definition, we are consistent with previous works on the same topic.

A power law dependency has also been analytically found by several theoretical works, for example by \citet{2017ApJ...847...29O} {and \citetalias{2020A&A...638A..52M}} for photoevaporation or by \citet{2019MNRAS.487...24G} for core-driven escape.
These works show that the slope of the valley is a good indicator for the dependence of the evaporation rate on the planets' distance from the host star and therefore the underlying evaporation mechanism.

It is important to specify that  {for the results obtained with the rectangular grid of initial conditions}, \(R_\mathrm{b}(p)\) is the lower boundary of the observed valley.
In contrast, observational studies,  {but also our theoretical results obtained for initial conditions derived from the observed Kepler planets}, report the middle of the valley.

{For the initial conditions derived from the Kepler survey, the absence of a completely empty gap (or a well-defined largest bare core as a function of period as in the rectangular grid) means that we cannot as simply quantify the valley position as for the rectangular grid. On the other hand, compared to the observed population, where various statistical methods must be used like support vector machines \citep[e.g.,][]{2018MNRAS.479.4786V} to determine the valley position and its slope, we are in the advantageous position to have a very large data set. We have therefore proceeded in the following simple way to determine the middle of the valley: We have binned the planets according to orbital period, with a bin width of 0.2 dex in \(\log{P}\), with partially overlapping bins at \(\log{P} = 0.6\), \(0.7\), \ldots \(1.7\), \(1.8\) (or \(1.9\) for the unbiased case). For each bin, we represent the radius distribution with a kernel density estimate, and get the position of the gap centre (local minimum in the radius distribution) from the zero point of the derivative. This procedure is similar to the one employed by \citet{petigura2022}}.

 {Finally}, to determine \(\tilde{R}_\mathrm{b}\) and \(\alpha \) from the simulations, we  {simply} make, as in \citetalias{2020A&A...638A..52M}, a least-square power law fit to the largest bare core radius (for the rectangular grid) respectively gap centre (for the Kepler initial conditions) as a function of orbital period at a given age, typically at \(5\,\mathrm{Gyr}\).

\begin{figure*} 
	\centering
    \includegraphics[width=\linewidth]{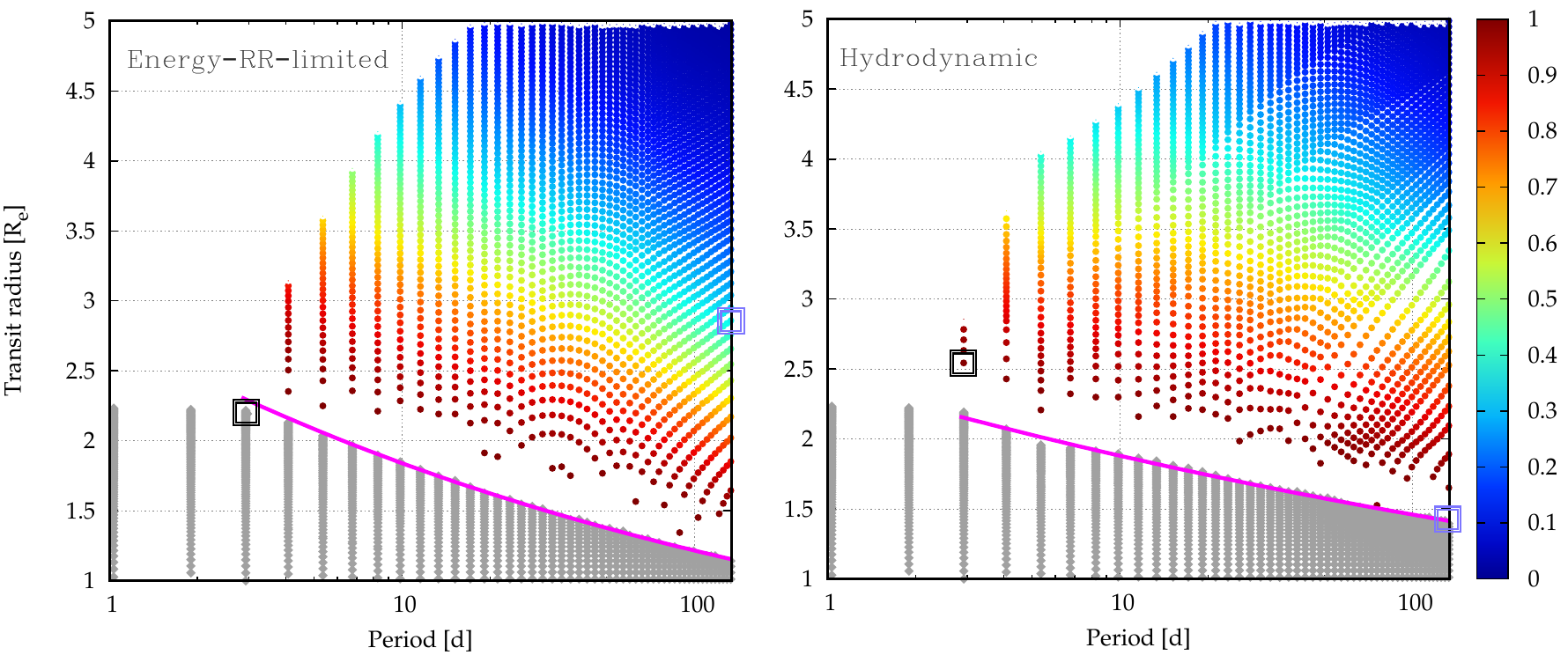}
    \caption{
        Result of planet evolution simulations in the orbital period--transit radius plane at \(5\,\mathrm{Gyr}\) (left panel: energy-/recombination-limited model; right panel: hydrodynamic model).
        Each dot represents a planet which is coloured based on the  {fraction of its initial H/He envelope mass that was evaporated}.
        Grey symbols indicate complete loss of the envelope, corresponding to sub-Neptunes that have evolved into super-Earths.
        The magenta lines show the fit to determine the slope \(\alpha \) of the valley as represented by the largest super-Earth at a given period (i.e.\@ the bottom of the valley).
        We note the shallower slope in the hydrodynamic model.
        The squares show two individual cases (one far planet in lilac, one close in black). Both start with identical initial conditions in the two models and are discussed in Sect.~\ref{subect:evoindividualcases}. They end up on different sides of the valley in the two models in a way to make the slope shallower in the hydrodynamic model.
    }\label{fig:grid}
\end{figure*}

\section{Results for the rectangular grid}\label{sec:results}
 {We present our planet evolution simulations by first examining the locus of the valley in the period--radius diagrams as predicted by the two evaporation models  {for the rectangular grid of initial conditions in Sect.~\ref{sec:grid}}. By performing case studies on individual planets in Sect.~\ref{subect:evoindividualcases}, we are then able to identify the cause of the distinct valley slopes. We then extend this analysis to the entire grid (Sect.~\ref{sect:compevapmodelsongrid}).} 

\subsection{Period-radius diagrams}\label{sec:grid}
\begin{table*} 
    \centering
    \caption{
        Valley radius at an orbital period of 10 days \(\tilde{R}_\mathrm{b}\) and valley slope \(\alpha \) as a function of orbital period \(p\) for both models and observational studies, quantifying the valley locus as \(\tilde{R}_\mathrm{b} \cdot {(p/\mathrm{10 \, days})}^\alpha \).  {We give the theoretical results found both for the rectangular grid of initial conditions (Sect.~\ref{sec:results}) and for the initial conditions derived from the Kepler survey results (Sect.~\ref{sect:keplercomp})}
    }\label{tab:slope}
    \begin{tabular}{p{12cm}@{\ }ll}
        \hline\hline
        & \(\tilde{R}_\mathrm{b}\) [\(R_\oplus \)] & Slope \(\alpha \) \\
        \hline
        Energy-/recombination-limited{, rectangular grid} \hrulefill{}
        & 1.84\tablefootmark{a}                   & \(-0.18\) \\
        Hydrodynamic{, rectangular grid} \hrulefill{}
        & 1.88\tablefootmark{a}                   & \(-0.11\) \\
         {Energy-/recombination-limited, Kepler initial conditions, unbiased} \hrulefill{}
        &  {2.39\tablefootmark{b} }                  &  {\(-\)0.18} \\
         {Hydrodynamic, Kepler initial conditions, unbiased }\hrulefill{}
        &  {2.28\tablefootmark{b}}                   &  {\(-\)0.11} \\
         {Energy-/recombination-limited, Kepler initial conditions, biased} \hrulefill{}
        &  {2.32\tablefootmark{b} }                  &  {\(-\)0.16} \\
         {Hydrodynamic, Kepler initial conditions, biased }\hrulefill{}
        &  {2.28\tablefootmark{b}}                   &  {\(-\)0.10} \\
        \citet{2018MNRAS.479.4786V} astroseismological sample \hrulefill{}
        & \(1.9 \pm 0.2\)\tablefootmark{b}        & \(-0.09^{+0.02}_{-0.04}\) \\
        \citet{2019ApJ...875...29M} CKS sample  {(identical \(\alpha \) also in \citealt{petigura2022})}\hrulefill{}
        & \(1.9 \pm 0.04\)\tablefootmark{b}       & \(-0.11_{-0.02}^{+0.02}\) \\
         {\citet{hovaneylen2023} high cadence sample} \hrulefill{}
        & \(1.84_{-0.07}^{+0.11}\)\tablefootmark{b}       & \(-0.096_{-0.027}^{+0.023 }\) \\
        \hline
    \end{tabular}
    \tablefoot{
        \tablefoottext{a}{Defined at the lower boundary of the valley.}
        \tablefoottext{b}{Defined in the centre of the valley.}
    }
\end{table*}

Figure~\ref{fig:grid} shows the simulated grid in the orbital period--transit radius plane for the two escape models at an age of \(5\,\mathrm{Gyr}\).
Clearly visible in both is the evaporation valley running diagonally downward, i.e.\@ the gap in radius between the super-Earth planets whose envelopes have fully evaporated and the sub-Neptunes which still have an envelope.
For the planets still possessing H/He, the colour code shows the fraction of the initial H/He envelope that has evaporated.
The closer to the valley, the higher this fraction, as expected.

With the divide being very distinct, a good fit of the slope can be achieved.
We note that there are some quasi-regular patterns emerging in the dots above the valley, and in the hydrodynamic model, some linear patterns are visible.
These patterns are simply the result from the regular grid, and, in the case of the hydrodynamic simulations, also consequences of the interpolation in the grid of tabulated evaporation rates.
For the hydrodynamic model, this also translates in the largest bare core as a function of period not being a completely smooth power law function, as it is the case for the energy-/recombination-limited model.
However, also for the hydrodynamic model, the simulations of the individual planets presented below exhibiting clear, physically understandable outcomes that are not dominated by interpolation artifacts, which, together with the small scales of the non-smoothness of \(R_\mathrm{b}\), mean that the interpolation does not significantly affect the quantities we are interested in (\(\tilde{R}_\mathrm{b}\) and \(\alpha \)).

In both models, there is a region with an absence of  {(sub\nobreakdash-)\allowbreak{}Neptunian} planets  {(i.e., planets with H/He)} at periods smaller than 2 or 3 days.
This corresponds to the evaporation (also called the sub-Neptunian) desert \citep{2016NatCo...711201L,2016A&A...589A..75M,2018A&A...620A.147B}.
It should be noted that the specific period marking the onset of the desert in our simulations shown here depends also on the fact that the most massive core we simulate has a mass of \(20\,M_\oplus \).
The minimum period is thus model dependent.
To quantify the valley locus, we only considered the planets having the smallest period for which there are still planets with an envelope.

The comparison of the two panels shows that while both models lead to a very similar position of the valley at 10 days \(\tilde{R}_\mathrm{b}\) of about \(1.8\) to \(1.9\,R_\oplus \), they differ in the slope, i.e.\@ in \(\alpha \) as is directly visually apparent.
In the energy-/recombination-limited model, the slope is clearly steeper than in the hydrodynamic model.
This is one of the key outcomes of this study.
In both panels, the magenta lines show the aforementioned power law fit to the largest, numerically found bare core as a function of period.
It makes the shallower slope in the hydrodynamic case even more apparent.

In Table~\ref{tab:slope}, we report the parameters of these fits, \(\tilde{R}_\mathrm{b}\) and \(\alpha \), together with the results of the observational studies of 
\citet{2018MNRAS.479.4786V}{,} \citet{2019ApJ...875...29M}{, and \citet{hovaneylen2023}}.
The data confirms the visual impression: the difference in \(\tilde{R}_\mathrm{b}\) (\(1.84\) and \(1.88\,R_\oplus \) in the energy-/recombination-limited and hydrodynamic model, respectively) is very small.
This difference is comparable to, or smaller than, the observational error bars, and thus not significant.

Our theoretical values for \(\tilde{R}_\mathrm{b}\) are for the bottom of the valley, while the two observational studies are for the middle.
Correcting for this difference would shift the theoretical \(\tilde{R}_\mathrm{b}\) to larger values by about \(0.2\)--\(0.3\,R_\oplus \) \citep{2020A&A...638A..52M,2021AJ....161..265D}, i.e.\@ to a radius between \(2.0\) and \(2.1\,R_\oplus \).
This is larger than the nominal observational values of about \(1.9\,R_\oplus \).
This could be an indication that both models overestimate escape.
A possible explanation for this could be that the interior structure and the escape models do not account for metals.
Because the planets considered here orbit late-type stars, metals do not cause much heating, but may lead to significant cooling, which would lower the escape \citep[e.g.,][]{2007P&SS...55.1426G}.
A metal-enriched instead of a pure H/He envelope would also affect the interior structure model via the equation of state and the opacities.
The impact of enriched envelopes was studied in \citetalias{2020A&A...638A..52M}.
It was found that a gas with a metal mass fraction of 10\% and 30\% would lead to a downward shift of the valley of about \(0.1\) and \(0.2\,R_\oplus \).
This would bring the theoretical predictions into better agreement with the observations. {Such enrichments could be a consequence of the formation process \citep{FortneyMordasini2013,BrouwersOrmel2020} or result from evolutionary magma-hydrogen interactions at the core-envelope interface \citep{MisenerSchlichting2023}. }

In contrast to \(\tilde{R}_\mathrm{b}\), the values of the slope \(\alpha \) differ clearly between the two theoretical models: in the energy-/recombination-limited model, a slope of \(0.18\) is found, while for the hydrodynamic model, the slope is \(0.11\).
This corresponds to a fractional difference of about 50\%, a difference that is clearly observationally relevant, given the error bars reported in the observational studie{s}.
The \(\alpha \) found in the updated numerical energy-/recombination-limited model used here is identical to the one derived analytically for a purely energy-limited model with a constant \(\eta_\text{XUV}\) derived in \citetalias{2020A&A...638A..52M}.
This shows that the introduction of an escape velocity dependent \(\eta_\text{XUV}\) does not significantly affect the results.

Comparing with the observationally inferred values of \(\alpha \) which are \(-0.09^{+0.02}_{-0.04}\) in \citet{2018MNRAS.479.4786V} and \(-0.11 \pm 0.02\) in  {both} \citet{2019ApJ...875...29M}  {and \citet{petigura2022}}, we see that the energy-/recombination-limited model yields with \(0.18\) a slope that is clearly too steep by more than two or three sigma.
The hydrodynamic model in contrast predicts a slope that is in excellent agreement with these observations.

The finding that a simple energy-limited model yields a steeper slope than a hydrodynamic model is not new: \citet{2017ApJ...847...29O} had already found an \(\alpha = -0.25\) for their simple energy-limited model with a constant efficiency factor, while their full hydrodynamic evaporation model \citep{2012MNRAS.425.2931O} without these assumptions yielded \(\alpha = -0.16\).
Both these values are, however, steeper than the observed values, in contrast to our new findings with the \citet{2018A&A...619A.151K} hydrodynamic model.
Another theoretical energy-/recombination-limited  {XUV photoevaporation} model \citep{2018MNRAS.479.5303L} found with \(\alpha = -0.15\) also a steeper slope than observed.
This makes it worthwhile to further investigate the result found here.

The \citet{2012MNRAS.425.2931O} model predates studies like \citet{2015A&A...576A..87S,2016ApJ...817..107O,2017A&A...598A..90F}, which found the boil-off phase as the first escape regime occurring after the dissipation of the disk.
Instead, their initial evaporation regime is X-ray driven, in contrast to our hydrodynamic model where boil-off is included.

 {We report here also} the masses of the largest bare core as a function of period in the two models, found like the radii with least-square power law fits.
This is an information of interest for radial velocity studies.
One finds for the energy-/recombination-limited model
\begin{equation}
    M_\mathrm{b,enRR}(p) = 9.6 \, {\left( \frac{p}{10\,\mathrm{days}} \right)}^{-0.64} M_\oplus
\end{equation}
and for the hydrodynamic model
\begin{equation}
    M_\mathrm{b,hyd}(p) = 10.6 \, {\left(\frac{p}{10\,\mathrm{days}}\right)}^{-0.41} M_\oplus.
\end{equation}
These values essentially reflect the mass--radius relation of silicate (MgSiO$_3$)-iron planets.
The analytical model of \citetalias{2020A&A...638A..52M} for an energy-limited model predicts for comparison that the exponent should be \(-0.66\).
This is very close to the value obtained numerically in the present paper (\(-0.64\)).
In the hydrodynamic model, the exponent is as expected significantly lower (\(-0.41\)).

\subsection{Evolution of specific cases}\label{subect:evoindividualcases}
 {The fact that \(\tilde{R}_\mathrm{b}\) is similar in the two models while the valley is clearly shallower in the hydrodynamic model means that the hydrodynamic model does not generate as large bare cores inside the 10-day period line as the energy-/recombination-limited model does. The opposite is true outside the 10-day period. This is quickly verified in Fig.~\ref{fig:grid}.}

 {Therefore, to understand the reasons for the different slopes, it is helpful to study two cases of individual planets. The first one is a far planet at a period of about 133 days, which remains a sub-Neptune in the energy-/recombination-limited model, but becomes a super-Earth in the hydrodynamic model (Sect.~\ref{sec:distant}).
The second one is a close planet at 3 days, which becomes a super-Earth in the energy-/recombination-limited model, but stays a sub-Neptune in the hydrodynamic model (Sect.~\ref{sec:closein}).
In other words, we study cases that end up on opposites sides of the valley in the two models in such a way that the slope is shallower in the hydrodynamic model. We select cases that are close/at the upper boundary of the super-Earths for the two distances.
In Fig.~\ref{fig:grid}, these individual planets are shown with squares (black for the close case, lilac for the far case).}

\subsubsection{Distant planet}\label{sec:distant}
\begin{figure*}
	\centering
    \includegraphics[width=\linewidth]{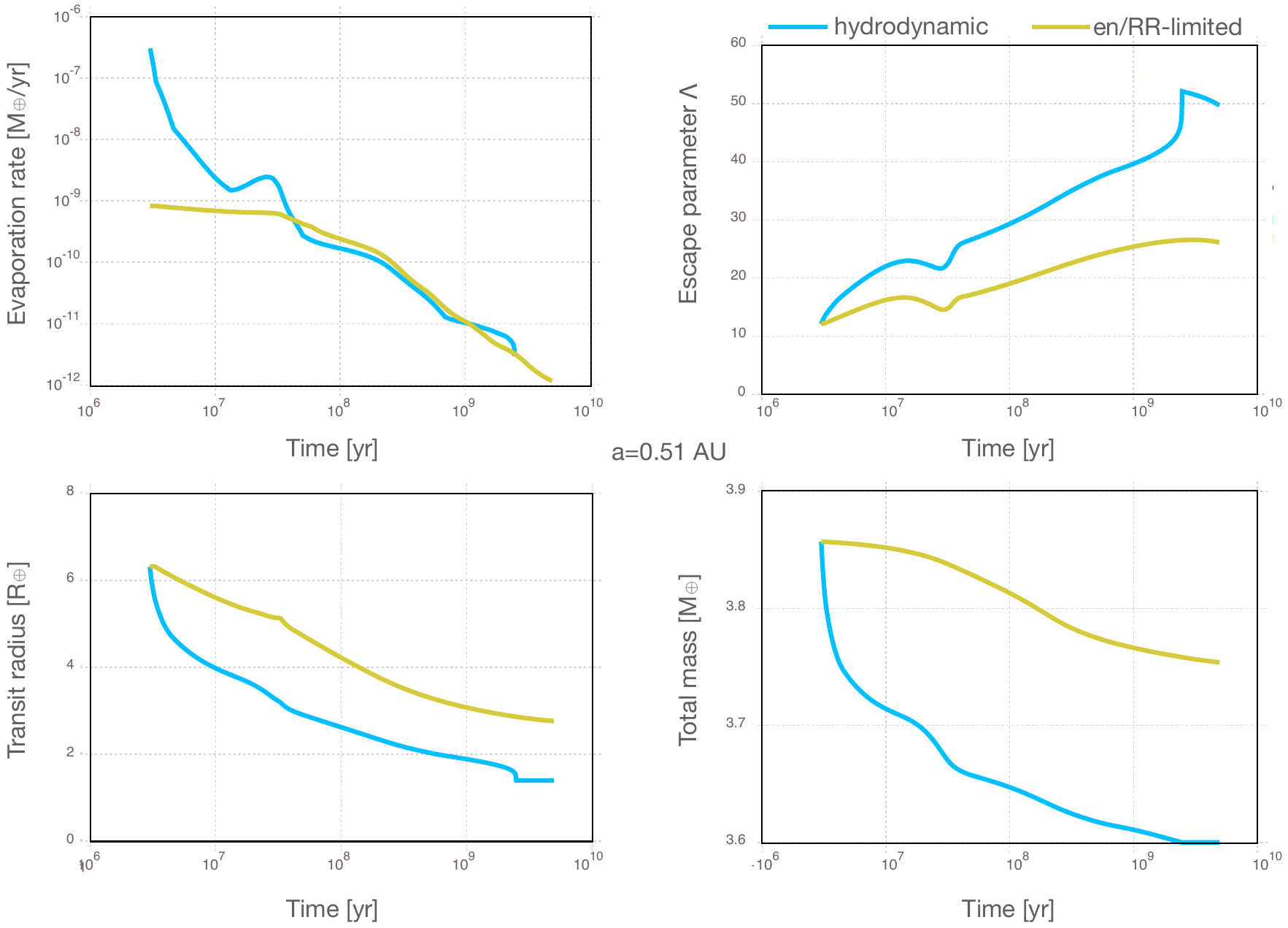}
    \caption{
        Temporal evolution of the evaporation rate (top left), restricted escape parameter \(\Lambda \) (top right), transit radius (bottom left), and total mass (bottom right) for a planet at \(0.51\,\mathrm{AU}\) and initial mass of approximately \(3.86\,M_\oplus \).
        The hydrodynamic and the energy-/recombination-limited model are shown.
        We note the very high mass loss in the hydrodynamic model in the initial boil-off phase.
        It is sufficient to eventually cause the total loss of the envelope.
        In the energy-/recombination-limited model which lacks this phase, the planet can in contrast keep a significant fraction of its envelope.
        The planet therefore resides in the end on different sides of the valley in the two models (super-Earth for the hydrodynamic model, sub-Neptune for the energy-/recombination-limited model). This is shown by the lilac squares in  {Fig.~\ref{fig:grid}}.
    }\label{fig:Far}
\end{figure*}
Figure~\ref{fig:Far} shows the temporal evolution of the evaporation rate, restricted Jeans parameter, transit radius, and total mass of a planet at an orbital distance of \(0.51\,\mathrm{AU}\) (period of 133 days) with an initial total mass of approximately \(3.86\,M_\oplus \).
The planet consists of a \(3.6\,M_\oplus \) silicate-iron core and a \(0.26\,M_\oplus \) H/He envelope.

The important result here is that in the hydrodynamic model, the planet completely loses its H/He envelope at about \(2.5\,\mathrm{Gyr}\), transforming the planet into a super-Earth, whereas in the energy-/recombination-limited model, the planet keeps about 60\% of the initial envelope till the end of the simulations and thus remains a sub-Neptune.
The final radii (at \(5\,\mathrm{Gyr}\)) in the two cases are about \(1.4\,R_\oplus \) and \(2.8\,R_\oplus \) (the former equal to the core radius), typical for planets below and above the evaporation valley.

The reason for this different evolution can be seen in the evaporation rate (top left panel in Fig.~\ref{fig:Far}).
We see that the hydrodynamic model initially predicts a much higher evaporation rate.
At the very beginning, the evaporation rate is more than 2 orders of magnitude higher.
It remains larger than the one in the energy-/recombination-limited model to about \(40\,\mathrm{Myr}\).
The reason for the higher evaporation rate can be seen in the Jeans-escape parameter (top right panel): we see that initially, \(\Lambda \) is about 12.
This puts the planet firmly into the boil-off regime \citep{2021A&A...650A..94K}, which leads to the very high escape rates in the hydrodynamic model.
During this time, the stellar continuum irradiation (mainly in VIS/IR), rather than the XUV irradiation, catalyses the escape.
 {A tell-tale sign of this is the local maximum in the escape rate in the hydrodynamic model at about \(30\,\mathrm{Myr}\), which is caused by the local maximum of the star's bolometric luminosity (see Fig.~\ref{fig:lumis}).
During boil-off,} the planet loses about half its H/He envelope in the first \(3\,\mathrm{Myr}\).
The escape gradually transitions into XUV-driven escape at about \(30\)--\(50\,\mathrm{Myr}\).
By this time, the radius of the planet has shrunk to a size that is comparable to one tenth of the Bondi radius (which is about \(40\,R_\oplus \)), as predicted by \citet{2016ApJ...817..107O}.
By this time, \(\Lambda \) has increased to about 30.
At later times, the hydrodynamic model predicts a comparable or somewhat smaller escape rate than the energy-/recombination-limited model, as can be seen in the top left panel.
However, by that time, the properties of the planets (namely the radii) have already diverged significantly between the two models, thus it is difficult to compare them.
We will do this later when we compare the escape rates at fixed planet properties.

To summarise, we see that for these distant low-mass planets, the hydrodynamic model predicts that comparatively more massive planets become super-Earth than in the energy-/recombination-limited, shifting the valley to larger radii because of boil-off.
This evaporation regime is included in the hydrodynamic model, but not {in} the energy-/recombination-limited one, which makes the difference.
This planet is actually a typical example of the first category of planets where the energy-limited approximation consistently fails by under-predicting the escape rate \citep{2021A&A...650A..94K}: planets with low-to-intermediate XUV irradiation and low gravitational potential.

\subsubsection{Close-in planet}\label{sec:closein}
\begin{figure*} 
	\centering
    \includegraphics[width=\linewidth]{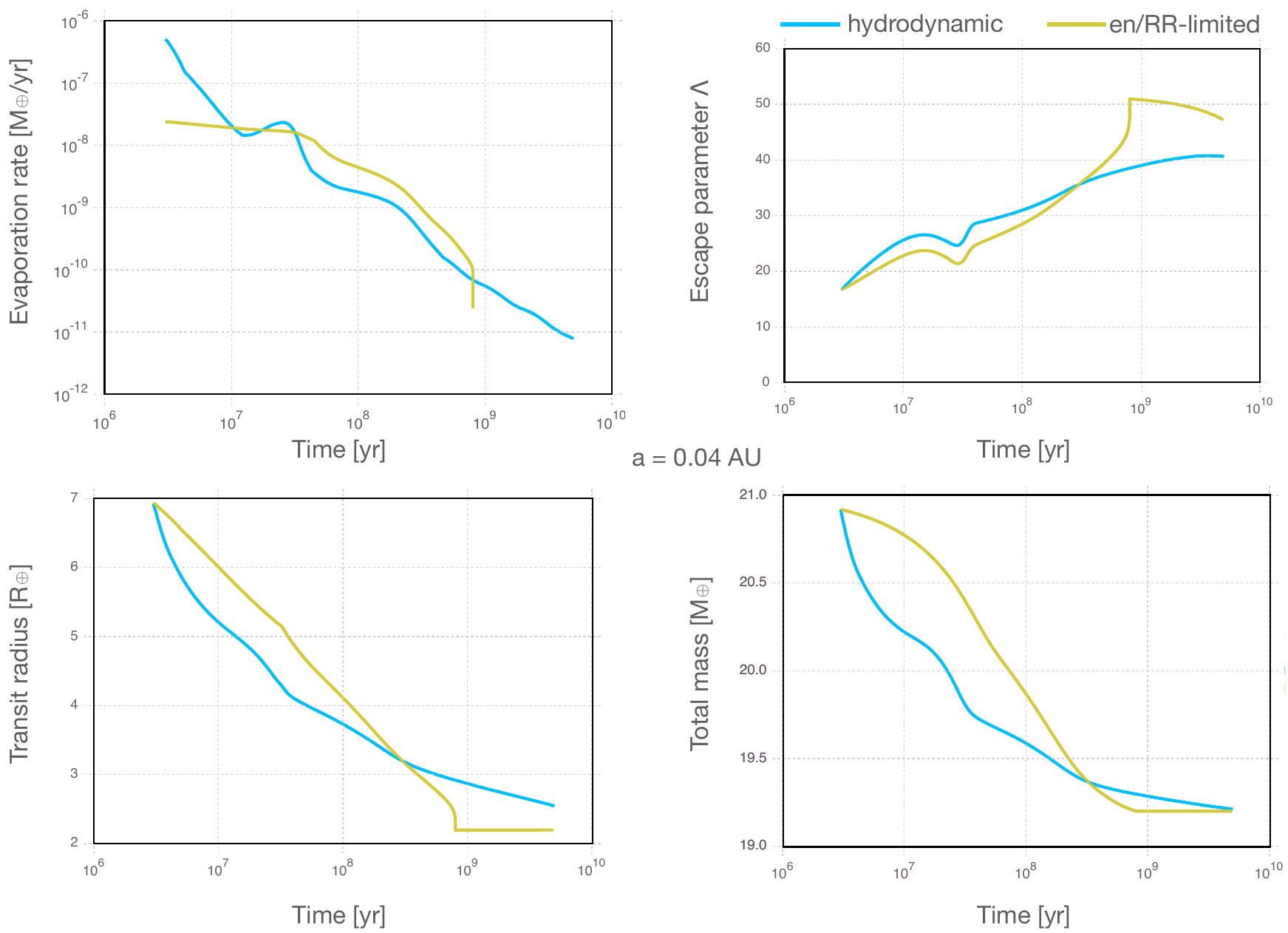}
    \caption{
        Temporal evolution of the evaporation rate (top left), restricted escape parameter \(\Lambda \) (top right), transit radius (bottom left), and total mass (bottom right) for a planet at \(0.04\,\mathrm{AU}\) and initial mass of approximately \(20.9\,M_\oplus \).
        The hydrodynamic and the energy-/recombination-limited model are shown.
        Despite the lack of the initial boil off phase in the energy-/recombination-limited model, the higher mass loss rate in this model after about \(30\,\mathrm{Myr}\) results in complete envelope loss, in contrast to the hydrodynamic model.
        The planets therefore reside in the end on different sides of the valley (super-Earth for the energy-/recombination-limited model, sub-Neptune for the hydrodynamic model). The underlying reason is the negligence of conduction in the energy-/recombination-limited model, leading to too high escape rates. {This planet is shown by the black squares in {Fig.~\ref{fig:grid}.}}
    }\label{fig:close}
\end{figure*}
The second case, a close-in massive planet, is shown in Fig.~\ref{fig:close}.
This is a planet at \(0.04\,\mathrm{AU}\) (orbital period of about 3 days), with an initial total mass of about \(20.9\,M_\oplus \).
The initial envelope mass is \(1.72\,M_\oplus \).
This may seem a small envelope mass for the significant core mass, but it is a consequence of the also very small orbital distance, that reduces in formation simulations the ability to accrete  {gas (see Eq.~\ref{eq:menve0}).}

Here, the key result is that the outcome is opposite to the  {distant case. 
As a matter of fact, in the energy-/recombination-limited model, the envelope is completely lost by \(800\,\mathrm{Myr}\).
Instead, in the hydrodynamic model, the planet keeps an envelope to the end of the simulation at \(5\,\mathrm{Gyr}\); see Fig.~\ref{fig:grid}.} 
 The remaining envelope mass at this time is actually tiny, but sufficient to lead to a radius of about 2.2 versus \(2.5\,R_\oplus \) for the two cases. 
In the case we study here, the difference of the two models is by construction small, since we have chosen a case that is just above/below the valley in the two models.

\begin{figure*} 
	\centering
    \includegraphics[width=\linewidth]{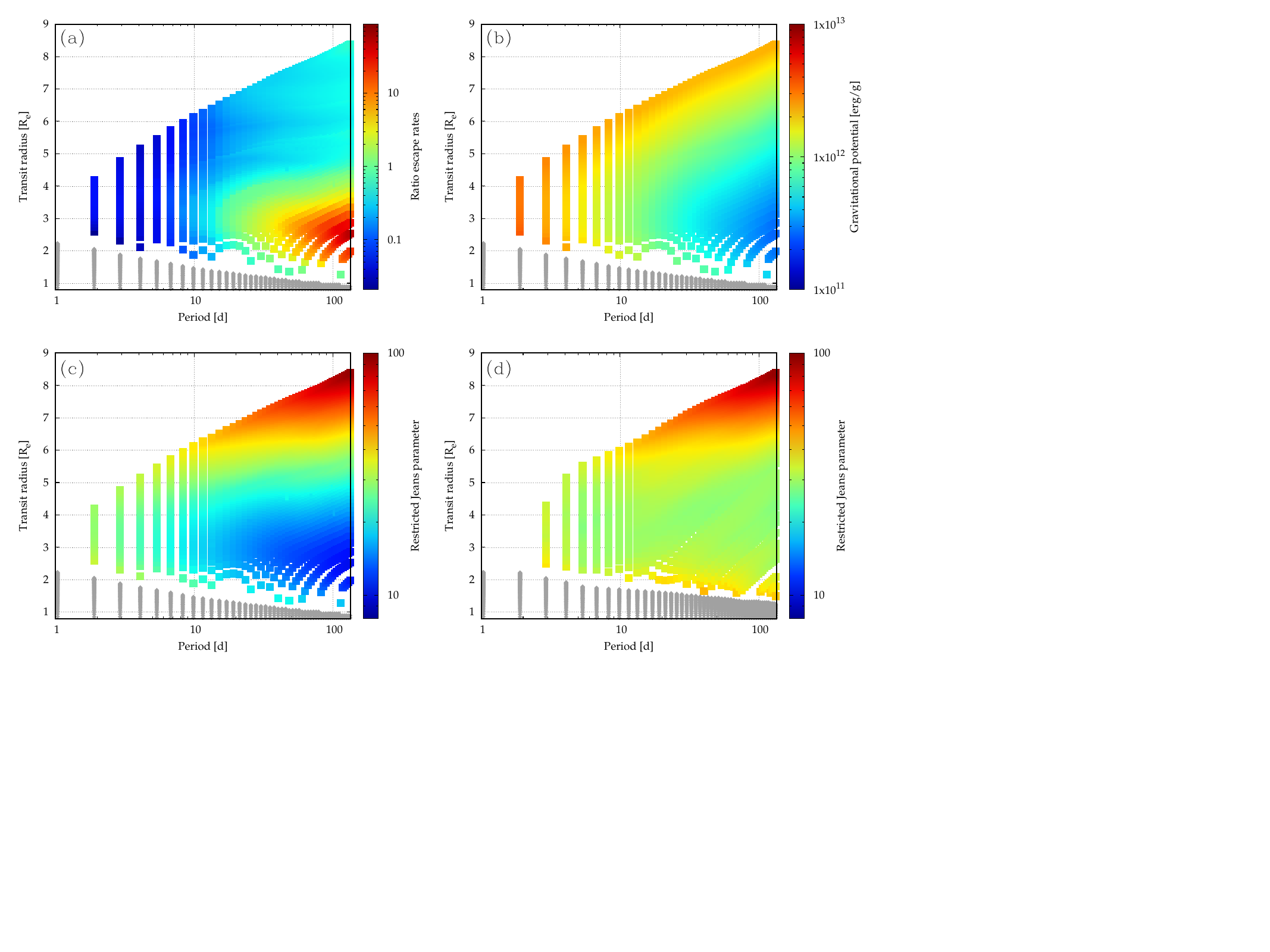}
    \caption{
        Comparison of the two escape models at \(50\,\mathrm{Myr}\) in the orbital period--transit radius plane.
        In Panels~(a), (b), and (c), the energy-/recombination-limited model is displayed while Panel~(d) displays the hydrodynamic model.
        In Panel~(a), the colour code shows the ratio of the instantaneous escape rates predicted by the two models, \(\dot{M}_\mathrm{hyd}/\dot{M}_\mathrm{enRR}\). One notes how the hydrodynamic model predicts higher escape rates for the distant small planets, but lower ones for the close-in planets. 
        Panel~(b) shows the gravitational potential of the planets.
        Panels~(c) and (d) colour code the restricted Jeans parameter \(\Lambda \). On the right, no \(\Lambda \lesssim30 \) occur, as the excess envelope has already boiled-off.
        Grey points are planets that have lost the entire H/He envelope.
    }\label{fig:compmodels}
\end{figure*}
 {The reason why the models lead to such different outcomes can be seen in the top left panel of Fig.~\ref{fig:compmodels}.
We see that as in the case of the distant planet,} the hydrodynamic model initially predicts a stronger mass loss, which is again due to boil-off.
However, compared to the far case, the boil-off is here less extreme, leading to a difference in the evaporation rates of initially a bit more than one order of magnitude.
This is due to the fact that as the planet already starts with a higher \(\Lambda \) of about 17.
Already after about \(10\,\mathrm{Myr}\), the evaporation rates become comparable in the two models, and after about \(30\,\mathrm{Myr}\), the escape rate in the energy-/recombination-limited model is consistently higher than in the hydrodynamic model.
Neither the radii nor the masses (dominated by the core mass anyway) of the planets differ strongly at this point. The similar \(M_\mathrm{p}\), \(R_\mathrm{p}\), and the identical \(T_\mathrm{eq}\) in the two simulations imply that \(\Lambda\) is also similar\footnote{It generally holds that decreasing \(R_\mathrm{p}\) implies decreasing the evaporation rate \(\dot{M}\) and increasing \(\Lambda\), while decreasing \(M_\mathrm{p}\) implies increasing \(\dot{M}\) and decreasing \(\Lambda\), so decreasing both \(M_\mathrm{p}\) and \(R_\mathrm{p}\) compensates to some extent (though the escape is more sensitive to variations in radii).}. Therefore, one can directly compare the evaporation rates in the two models. The lower rate in the hydrodynamic model is thus not merely a consequence of different planet properties, but a genuine consequence of the more complex physics included in the hydrodynamic model, and more specifically in the different temperature structure compared to that assumed in the energy-limited approximation, as we shall further discuss in the following section.
The difference in the predicted escape rates is not very large (factor 2 to 3), but integrated over time, this is sufficient for complete evaporation in the energy-/recombination-limited, but not the hydrodynamic model.

The underlying reason is that this planet is a typical example of the second regime where the energy-limited approach consistently fails by over-predicting the escape rate \citep{2021A&A...650A..94K}, namely planets characterised by high XUV irradiation and high gravitational potential.

\subsection{Comparison of the escape models on the entire grid}\label{sect:compevapmodelsongrid}

The previous section demonstrated the different outcomes of two selected planets using the two escape models, and which effects played a significant role.
We now generalise this comparison to the entire grid of planets we simulated \citep[for another systematic comparison, see also][]{2021A&A...650A..94K}.

Figure~\ref{fig:compmodels} compares the two models at an age of \(50\,\mathrm{Myr}\).
At this time, boil-off in the hydrodynamic model will already have ceased, but strong XUV-driven evaporation (because we are still at early times) will be ongoing.
Four panels are shown in the period--radius plane, colour-coding different quantities.
 {In Panels~(a), (b), and (c), the results of the energy-/recombination-limited model are shown with coloured dots, in Panel~(d) they show the results of the hydrodynamic model.}
In all cases we see the valley, which is, however, not yet at the same position as in Fig.~\ref{fig:grid}, which shows the situation at \(5\,\mathrm{Gyr}\).

Panel (a) shows  {colour-coded} the ratio of the escape rate predicted by the hydrodynamic model over the escape rate predicted by the energy-/recombination-limited model, \(\dot{M}_\mathrm{hyd} / \dot{M}_\mathrm{enRR}\).
The latter is the rate that is actually used to model the evolution of the planets shown in the panel.
The former is merely calculated given the properties of the planets at this moment, but is not used for the evolution.
This allows to compare the two models at fixed planet properties which was not easily possible in the analysis of the two individual cases.

 {The plot reveals the two aforementioned shortcomings of the energy-/recombination-limited approximation. The first regime is shown by the blue points: for close-in compact and massive planets with high gravitational potential exposed to high XUV irradiation, the energy-/recombination-limited model overestimates the escape rate relative to the hydrodynamic model. This is shown in Panel~(b), which colour codes the gravitational potential for the energy-/recombination-limited case.
The reason for the incorrect results of the energy-limited approximation in this particular regime has been described in \citet[][their Sect.~5.2]{2021A&A...650A..94K}: the assumed thermospheric temperatures underlying the energy-limited approximation are much higher than the ones found when directly solving the governing equations in the hydrodynamic model.
The discrepancy is a consequence of the lack of downward heat conduction underlying the energy-limited approximation, leading to excessively high temperatures and thus loss rates.
In reality, in the dense atmospheres at the XUV absorption height of planets with high gravitational potential, thermal conduction is a significant process, leading to lower temperatures and thus escape rates.}

 {The second regime is shown by red and yellow point{s}: for planets with low-to-intermediate XUV irradiation and low gravitational potential, the energy-/recombination-limited model underestimates the escape rate relative to the hydrodynamic one. Here, the energy-limited approximation fails again because of the incorrect assumed temperature structure \citep{2021A&A...650A..94K}: for such planets, boil-off is the dominant escape mechanism.
However, the energy-limited approximation implicitly neglects thermal energy already available in the atmosphere resulting from optical and infrared stellar irradiation.
When \(\Lambda \) is low (less than about 30), this thermal energy is comparable to the binding energy, leading to boil-off.
It is more intense for planets with lower masses, while planets more massive than approximately \(10\,M_\oplus \) are less affected \citep{2016ApJ...817..107O}.
The impact of boil-off is illustrated by Panels~(c) and (d).}
 {In Panel~(c) we see that the regime where the hydrodynamic model predicts significantly higher escape rates corresponds to the planets with the lowest \(\Lambda \) values.
For them, boil-off and rapid mass loss would occur in the hydrodynamic model, but this is neglected in the energy-/recombination-limited approximation.
This strong mass loss rapidly reduces the planet radius, increasing \(\Lambda \) until it approaches about 30, when boil-off stops.
This explains what is seen in Panel~(d): here, in the hydrodynamic model, there are no planets with \(\Lambda \) less than about 30, because the excess envelope mass has already boiled-off by \(50\,\mathrm{Myr}\) (or even much faster, as we saw for the two individual cases studied above).}

 {Apart from these two regimes of discrepancies, there is also a significant part where the two models yield similar escape rate (light blue - cyan - green colours in panel a).
The discrepant regimes, are, however, the ones setting the valley slope.}

\section{Results for initial conditions derived from Kepler observations}\label{sect:keplercomp}
\begin{figure*}
	\centering
    \includegraphics[width=0.8\linewidth]{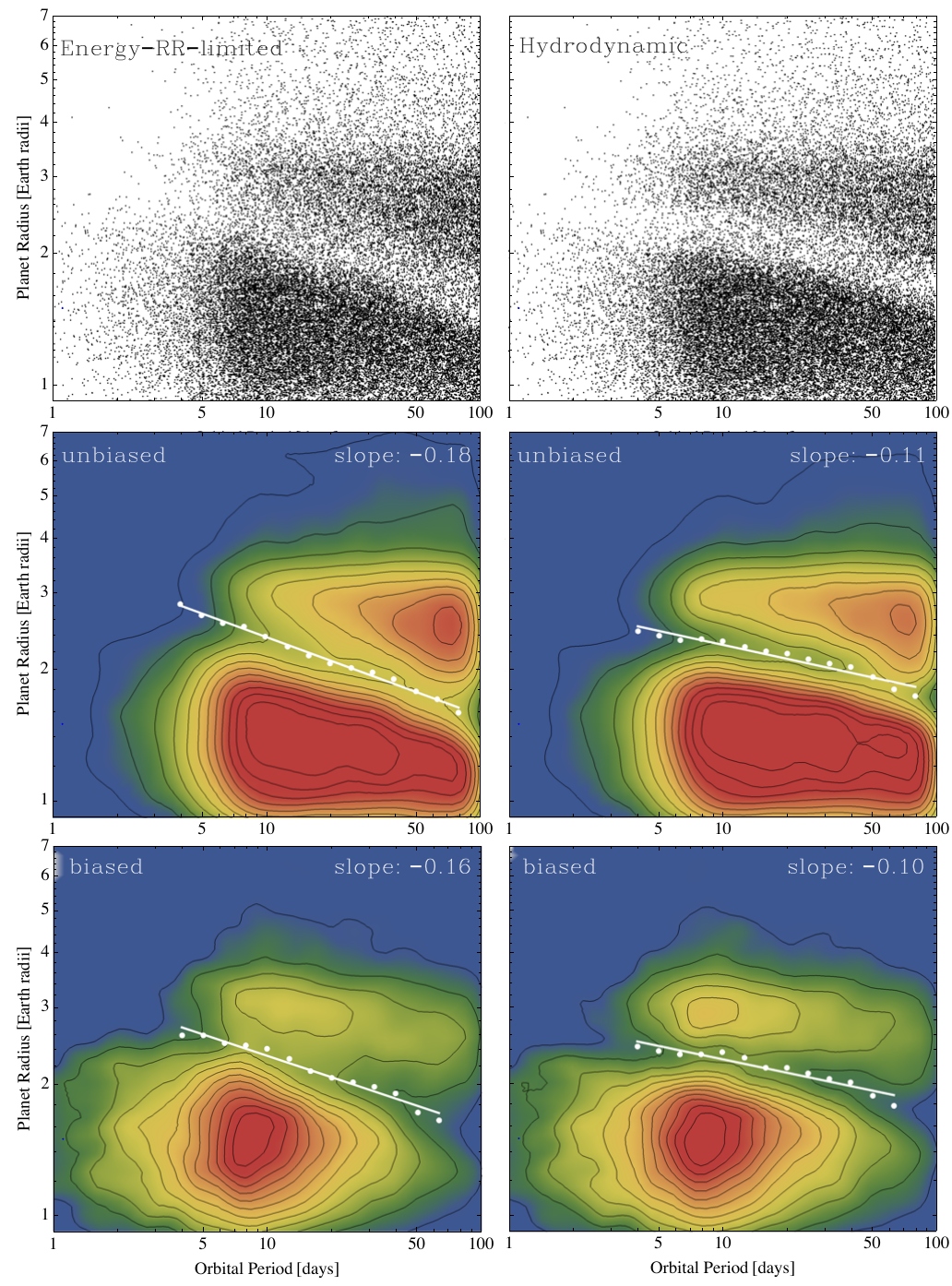}
    \caption{ {Transit radius as a function of orbital period at \(5\,\mathrm{Gyr}\) for distributions of the initial envelope mass, core mass, and orbital period derived from Kepler observations \citep{rogersowen2021}. Left column: energy/radiation-recombination-limited escape model. Right column: hydrodynamic escape model. Top row: raw scatter plot of the unbiased synthetic populations. Middle row: 2D Gaussian Kernel Density Estimation of the unbiased synthetic populations. Bottom row: as in the middle, but after applying a detection bias representative of the Kepler survey, which disfavours small distant planets. White dots and lines indicate the valley position. One notes in all cases the shallower slope in the hydrodynamic model. }
    }\label{fig:prkep}
\end{figure*}
\begin{figure}
	\centering
    \includegraphics[width=0.9\linewidth]{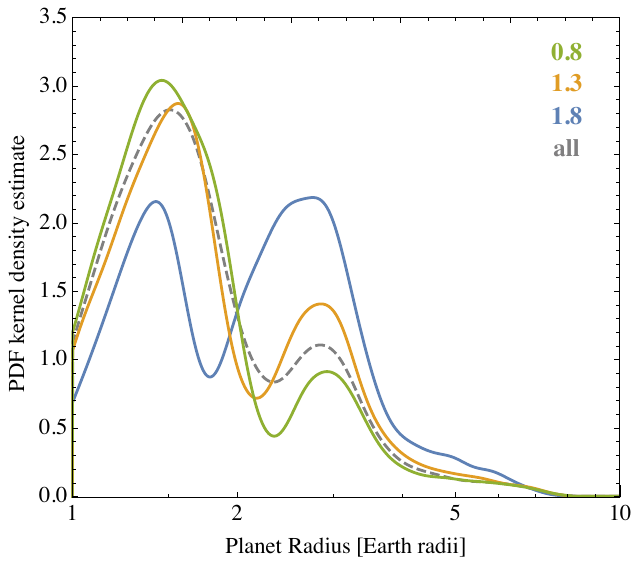}
    \caption{ {Kernel density estimate of the distribution of the radii in the biased synthetic population obtained with the hydrodynamic escape model. The grey dashed line includes all detectable planet (at all orbital periods). The green, orange, and blue lines include planets with \(\log \) (period/day) of \(0.8 \pm 0.1\), \(1.3 \pm 0.1\), and \(1.8 \pm 0.1\). One sees how the centre of the valley shifts to smaller radii with increasing orbital period.}  
    }\label{fig:histoR5biaed}
\end{figure}
 {The goal of this part is to understand if the main result obtained with the rectangular grid of initial conditions --- the distinct slopes of the valley --- also persist if we use more complex (and  more realistic) distributions of the initial conditions, and apply a synthetic detection bias of the Kepler satellite survey to the model output. For this, we analyse the synthetic planet population obtained with the initial conditions described in Sect.~\ref{Sect:compKeplerinit}, considering both the biased and the unbiased data.}

 {The top row of Fig.~\ref{fig:prkep} shows the scatter plot of transit radius as a function of orbital period for the two evaporation models at \(5\,\mathrm{Gyr}\). No detection bias was applied. Compared to the equivalent plot for the rectangular grid of initial conditions (Fig.~\ref{fig:grid}), we see a number of similarities, but also differences. Similarities are the scarcity of close-in planets with large radii in the top left corner of the plots (the evaporation desert), and the presence of the evaporation valley running diagonally downward. An important similarity regarding the main subject of the paper is that the slope of the valley is steeper for the energy/recombination-limited model than for the hydrodynamic model. We quantify the slopes in the following. We also see the following differences: the period distribution is by construction different, with an increase in planet frequency from 1 to about 8 days of period, followed by a distribution constant in \(\log{P}\). This simply reflects the initial conditions (Eq.~\ref{eq:perioddist}). A more important difference concerns the presence of an overdensity of planets in the region immediately above the valley. This populates the sub-Neptunian peak in the radius histogram. At even larger radii (\(\gtrsim 3.5\,R_\oplus \)) the frequency of planets drops strongly (the cliff, \citealt{Kite_2019}). Both these features are important aspects of the observed planet distribution (e.g., \citealt{Petigura_2020}), but were absent in the rectangular grid. We see that with the inferred core and envelope mass distribution of \citet{rogersowen2021}, we find these features also with our different forward (escape and interior structure) model. As a last difference we see that the valley is not fully empty, but contains some planets. There are two types of planets in the valley: First, massive bare cores (\(\gtrsim 20-30 \, M_\oplus \)) that started with very small post-formation envelopes (\(0.01 \, M_\oplus \) or less), such that they were fully evaporated despite the large core mass. These planets populate the gap from `below' and are dominant in the lower half of the depleted gap area. Second, lower mass planets that are in the process of losing the final part of their envelope. They only contain less than \(\mathbin{\sim} 1\%\) of their initial envelope mass. These planets populate the gap from `above' and dominate in the upper half of the gap.}

 {Our procedure to obtain the gap locus (centre) was described in Sect.~\ref{sect:quantifyingvalleylocus}. The approach employing a running mean is illustrated with Fig.~\ref{fig:histoR5biaed}. It shows the Kernel Density Estimate of the radius distribution for the hydrodynamic model, including all detectable planets (grey dashed line) as well as planets within three different period intervals. We see how the valley position systematically moves to smaller radii with increasing orbital distance.}

 {From the minima in the distributions, we obtain 14 (13) positions for the middle of the valley, for the unbiased and biased population, respectively. These positions are shown with dots in the middle and bottom panels of Fig.~\ref{fig:prkep}. These panels show 2D Gaussian Kernel Density Estimations of the unbiased 
(middle) and biased (bottom) population. The impact of the detection bias which removes distant small planets is clearly visible. The two types of planets (super-Earths and sub-Neptunes) and the cliff are also clearly visible in this representation.}

 {Finally, we have made least-square fits to these points. They are shown with white lines in the figure. For the unbiased case, we find slopes of \(\alpha = -0.18\) and \(-0.11\) for the energy-/recombination-limited and the hydrodynamic model, respectively. These values are identical to those derived for the rectangular grid (Table~\ref{tab:slope}). We thus find that using these very different and more realistic initial conditions does not affect the main result found with the idealised rectangular grid of initial conditions. This indicates that the imprint of the different evaporation models is quite solid, and not strongly dependent on the initial conditions, like for example the assumed post-formation envelope mass.}

 {In the biased case, which is the one most directly comparable with observations, we find slopes of \(-0.16\) and \(-0.10\) for the energy-/recombination-limited and the hydrodynamic model, respectively. Applying the detection bias thus makes the slopes slightly less steep for both cases, an effect that should be kept in mind when comparing (unbiased) model predictions and observations, although the difference is tiny (see also \citealt{petigura2022}). More importantly, however, these values still compare in the same way to the observed values as was already found with the rectangular grid: the slope found with the hydrodynamic model is in very good agreement with the observed slope (covering a 1--\(\sigma \) range of about \(-0.13\) to \(-0.07\) depending on reference), whereas the energy-/recombination-limited model yields a too steep slope. Thus, applying a detection bias does also not alter the main conclusion of the study based on the idealised rectangular grid.}

 {Regarding the absolute position of the valley at 10 days period, we find that the middle of the gap is predicted to be at about \(2.3\,R_\oplus\) for both evaporation models (see Table~\ref{tab:slope}). As for the rectangular grid, these are larger radii than observed (\(1.9 \pm 0.2 \, R_\oplus\) according to \citealt{2018MNRAS.479.4786V}). Thus, our theoretical model seems to overestimate in a general way the strength of evaporation. As already discussed in Sect.~\ref{sec:grid}, the presence of a lot of metals as coolants in the atmospheres might explain the difference. In \citetalias{2020A&A...638A..52M} it was found that a metal mass fraction of about \(Z=0.5\) would shift the valley downward by approximately \(0.4\,R_\oplus\). This calculation did, however, employ a highly uncertain scaling of the escape rate with \(Z\) derived from photoevaporation models of protoplanetary disks \citep{ercolano2010}. Systematic tabulations of atmospheric escape rates found with hydrodynamic models as the one used here but now as a function of \(Z\) (represented, e.g., as scaled solar composition) including very high values instead of pure hydrogen would help to further clarify this point. Observationally, future measurements of the atmospheric composition of sub-Neptunes with JWST will also be useful for a better understanding.}

 {In the radius histogram including all detectable planets obtained from the model (dashed line in Fig.~\ref{fig:histoR5biaed}), the super-Earth peak is almost three times as high as the sub-Neptune peak. Observationally, they are in contrast of similar height \citep{fultonpetigura2018,ZhuDong2021}. It is not surprising that we get such a discrepancy, because the initial condition distributions we use were derived from an inference analysis utilising another evaporation (forward) model. Modifying the initial conditions would, however, allow to change this ratio: by shifting the core mass distribution to more massive values, a higher fraction of planets would be massive enough to keep a H/He envelope and populate the sub-Neptunian peak. The minimum core mass necessary to keep some H/He at a given orbital distance was analytically derived in \citetalias{2020A&A...638A..52M} (their Eq. 29).}

\section{Summary and conclusions}\label{sec:discussion}

In this work, we tested both a simpler XUV-driven energy-/recombination-limited escape model and a complex hydrodynamic escape model \citep{2018A&A...619A.151K} against  {a key observational constraint}, the valley slope. The latter model includes the boil-off, blow-off, and Jeans escape regimes. The comparison was done by coupling the two escape models to a model for the temporal evolution of planetary interiors. This  {interior} model solves the classical spherically symmetric interior structure equations. With these models, we simulated the evolution of 6000 planets on an  {idealised rectangular} grid in orbital period and mass{, and for about 37\,000 planets with initial conditions (period, core, and envelope mass) derived from an inference analysis of the Kepler survey planet population \citep{rogersowen2021}.} We studied the valley locus predicted by the two escape models at \(5\,\mathrm{Gyr}\). 

We find that the hydrodynamic model leads to a valley slope \(\mathrm{d} \log{R} / \mathrm{d} \log{p} = -0.11\)  {both for the rectangular grid and the unbiased synthetic Kepler planet population. Applying a simple detection bias of the Kepler survey \citep{petiguramarcy2018} leads for the hydrodynamic model to a slope of \(-0.10\). These slopes} agree closely with the observational result derived by \citet{2018MNRAS.479.4786V} (\(-0.09^{+0.02}_{-0.04}\)) and  {\citet{2019ApJ...875...29M,petigura2022}} (\(-0.11 \pm 0.02\)). As past photoevaporation models, the simple energy-/recombination-limited escape model in contrast predicts a too steep slope of \(-0.18\)  {for the rectangular grid and the unbiased synthetic Kepler population, and of \(-0.16\) for the biased synthetic population}. Regarding the radius of the lower boundary of the valley at a fixed 10-day orbital period, both models yield similar values  {for the rectangular grid}, namely about \(1.8\)--\(1.9\,R_\oplus \).

The too steep a slope in the energy-/recombination-limited escape model is caused by two limitations of this approximation \citep{2021A&A...650A..94K}, as is found by comparing the escape rates for both individual planets and the entire grid: In particular, it underestimates escape rates for distant, fluffy low-mass planets while simultaneously overestimating it for close-in, compact massive planets. The former is caused by the omission of the boil-off escape regime in the purely XUV-driven energy-/recombination-limited model, while the latter can be explained by its negligence of heat conduction in the atmosphere.

Boil-off \citep{2015A&A...576A..87S,2016ApJ...817..107O,2017A&A...598A..90F} causes a rapid mass decrease in the first few megayears for fluffy planets with a low restricted Jeans-escape parameter. These are planets with considerable thermal energy stored in their atmosphere relative to their gravitational potential. This initial mass loss is significant enough to alter the slope of the valley by evaporating the atmosphere of more massive planets at larger distances. It is interesting to note that the escape rate in the boil-off regime depends via the planetary equilibrium temperature on the stellar continuum irradiation (VIS/IR), and not the XUV irradiation. This is a property it shares with core-driven escape, contrasting the purely XUV-driven energy-/recombination-limited model. Our results suggest that a combination of aspects of both models (namely heating both in VIS/IR and XUV) yield a valley slope agreeing with observations. 

The second limitation, the negligence of heat conduction in the energy-/radiation-limited approximation produces higher temperatures in the atmosphere than when conduction is cooling the upper atmosphere, as it is the case in the hydrodynamic model. The energy-/radiation-limited model therefore overestimates the temperature, leading to an excessive mass loss rate. This effect is prevalent for massive close-in planets, which are highly  {XUV irradiated} and feature compact atmospheres \citep{2021A&A...650A..94K}. By including conduction-cooling, the hydrodynamic model predicts lower mass loss rates over time for such planets, leaving lower mass planets still with a H/He envelope, which also alters the valley slope. In combination, the two shortcomings act together in the same direction: the too weak evaporation at larger distances (resulting in smaller evaporated cores) and too strong evaporation at smaller distances (resulting in larger evaporated cores) give the valley a too steep slope in the energy-/recombination-limited model.

Our results indicate that the more realistic description of the thermospheric temperature structure in the hydrodynamic model relative to the energy-/recombination-limited model is critical. It allows to reproduce one of the most important observational constrains for escape models, the valley slope. 

Future work will address the evaporation valley's dependency on host star mass  {\citep[see also][]{gupta2022}} and the effect of including metals, which may act as coolants. When compared with observational studies exploring the temporal  {\citep{2020AJ....160..108B,2021ApJ...911..117S,2021AJ....161..265D,petigura2022,hovaneylen2023}} and stellar mass dependency  {\citep{fultonpetigura2018,2019ApJ...874...91W,berger2020,petigura2022,dichang2022}}, this should allow {one} to develop an even better understanding of the origin of the radius valley.

\begin{acknowledgements}
 {L.A., A.V.O., and C.M. acknowledge support from the Swiss National Science Foundation under grant 200021\_204847 `PlanetsInTime'. Parts of this work have been carried out within the framework of the NCCR PlanetS supported by the Swiss National Science Foundation under grants 51NF40\_182901 and 51NF40\_205606. Calculations were performed on the Horus cluster at the University of Bern. The research described in this paper was carried out in part at the Jet Propulsion Laboratory, California Institute of Technology, under a contract with the National Aeronautics Space Administration.}
\end{acknowledgements}

\bibliography{ads}

\appendix

\section{ {Impact of the post-formation luminosity}}\label{appendix:lumi}
 {In this appendix, we evaluate the sensitivity of our results for the valley locus on the post-formation luminosity \(L_0\) or a closely related quantity, the specific entropy \(s_0\) in the inner convective zone.}

\subsection{ {Parameterization and expected range of \(s_0\)}}
 {As in \citetalias{2020A&A...638A..52M}, our nominal approach is to assume an initial luminosity \(L_0\) that is given by a fit to results of planet formation population syntheses \citep{2014A&A...566A.141M},}
\begin{equation}\label{eq:l0par}
L_0/L_\mathrm{\jupiter}\approx 0.008 \times {\left(\frac{M_\text{core}}{M_\oplus}\right)}^{2.5}.
\end{equation}
 {In this equation, \(L_\mathrm{\jupiter}\) is the intrinsic luminosity of Jupiter today (about \(8.7\times 10^{-10} L_\odot\)). It is clear that in reality, depending on the particular formation history of a planet, the post-formation luminosity may vary \citep[e.g.,][]{Marley2007,Mordasini2017,Cumming2018,Marleau2019}. \citet{Mordasini2017} for example found a spread in post-formation entropy \(s_0\) in the planet mass range of interest here of about 1 to \(1.5\,k_\mathrm{B}/\text{baryon}\) at fixed envelope mass.}

 {These earlier works investigated the post-formation entropy mainly in the context of giant planets and their detectability with direct imaging. More recently, the impact of the post-formation thermodynamic state was also addressed for evaporating planets: \citet{Owen2020} showed how the post-formation entropies of young evaporating planets might be constrained by observations. \citet{Kubyshkina2021} found that the initial entropies of planets have a minor or even absent effect on most of the population of evolved planets with ages of \(\mathbin{\sim} 1\,\mathrm{Gyr}\). Only for planets suffering extremely strong atmospheric mass loss, \(s_0\) was found to be of importance. A low importance of the entropy is also expected from the rather weak dependency of the thickness of the convective zone of H/He envelopes on the age and thus entropy \citep{Lopez2014}. }

 {Our approach here is to re-run the evolutionary simulations on the rectangular grid of initial conditions with the two evaporation models, but instead of using Eq.~\ref{eq:l0par}, we cover a wide range of \(s_0\), including the one suggested by more modern formation models. We can then systematically study how \(s_0\) affects the valley location. For this, we generalise the parameterisation of \(s_0\) of \citet{Malsky2020},
\begin{equation}\label{eq:s0par}
s_0 = s_\mathrm{0,n} + \frac{M_\mathrm{p}}{25\,M_\oplus} k_\mathrm{B}/\text{baryon}.
\end{equation}
\citet{Malsky2020} fixed \(s_\mathrm{0,n}\) to \(6\,k_\mathrm{B}/\text{baryon}\). Here, we generalise this and use \(s_\mathrm{0,n} = 6\), \(7\), \(8\), and \(9\,k_\mathrm{B}/\text{baryon}\).}

\begin{figure*}
	\centering
    \includegraphics[width=\linewidth]{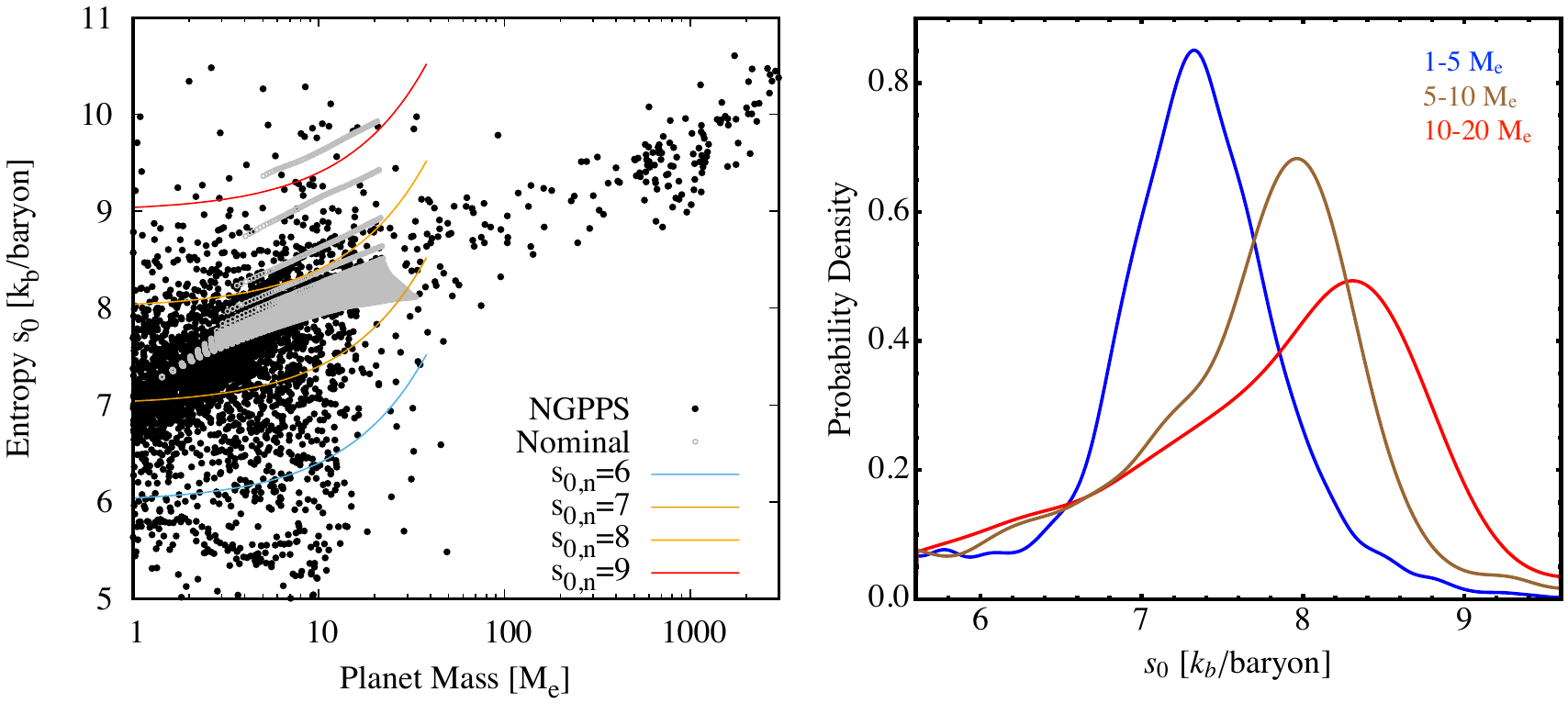}
    \caption{ {Left panel: post-formation entropy \(s_0\) as a function of planet mass. The black dots show \(s_0\) predicted by core accretion planet formation simulations in the New Generation Planet Population Synthesis NGPPS \citep{Emsenhuber2021b}. Gray dots are the \(s_0\) in the nominal case (Eq.~\ref{eq:l0par}). The coloured lines finally represent the four parameterisations used in this appedix, which are a generalisation of \citet{Malsky2020}. Right panel: kernel density estimation of the distributions of \(s_0\) in the NGPPS of three intervalls of the total planet mass. One sees how the mode of the distribution shifts to higher values with increasing planet mass.}}\label{fig:initialentro}
\end{figure*}

 {Before discussing the results of these grid simulations with different \(s_\mathrm{0,n}\), we compare the \(s_0\) obtained in this way with the ones predicted by the recent comprehensive planet population synthesis simulation NGPPS \citep{Emsenhuber2021b}. These simulations represent a much improved update to the ones used to derive the original fitting equation \citep{2014A&A...566A.141M}. These NGPPS results are generated with the Generation III Bern Model \citep{Emsenhuber2021a} which is a complex end-to-end formation and evolution model based on the core accretion paradigm. The model solves the 1D internal structure equations during the formation (both attached and detached state), and evolutionary phases. In the luminosity calculation it takes into account the accretion of planetesimals and gas, the cooling and contraction of the envelope, radiogenic heating, as well as giant impacts. The model also takes the concurrent formation of several planets in one disk into account, in contrast to the older \citet{2014A&A...566A.141M} syntheses, which used the one-embryo-per-disk approximation. This leads to more diverse formation pathways \citep{Emsenhuber2021b}.}

 {The left panel of Fig.~\ref{fig:initialentro} shows as black dots the entropy at the core-envelope boundary of the planets the NGPPS simulation. The nominal synthetic population NG76 is shown at the moment when the the gas disk dissipates, which corresponds to ages between \(1\) and \(10\,\mathrm{Myr}\). The host star mass is \(1\,M_\oplus\). We see that generally speaking, the (mean) entropy is an increasing function of the planet mass. Especially at smaller masses, there is significant spread in \(s_0\). Given the high density of points in this mass range, it is however difficult to get a quantitative picture of the distribution of \(s_0\) from the scatter plot. Thus, in the right panel we additionally show the kernel density estimation of the distribution of \(s_0\) for three mass ranges of interest for our study. We see that the mode indeed increases with mass, lying at about \(7.3\), \(8.0\), and \(8.3\,k_\mathrm{B}/\text{baryon}\) for the low, mid, and high mass range. The FWHM is about \(1\) to \(1.5\,k_\mathrm{B}/\text{baryon}\) in the three cases. This spread is thus similar to the one in the older syntheses.}

 {The grey points show the \(s_0\) obtained with the nominal approach (Eq.~\ref{eq:l0par}). Since we here specify \(L_0\) and not directly the entropy, and given that the atmospheric boundary conditions (in particular the temperature) also affect the relation between \(L_0\) and \(s_0\) (see \citealt{Marleau2014} and \citealt{Kubyshkina2021}), a range of \(s_0\) occur. The points fall on lines of fixed orbital distance (or equilibrium temperature), with the high \(s_0\) values corresponding to the closest distances. We also see that the majority of the grey points falls into a similar range as also the majority of the black points do. Thus, the simple fit derived from the older population synthesis still seem to capture --- at least in a rough way --- the new NGPPS results for \(s_0\). The four coloured lines finally show Eq.~\ref{eq:s0par} with the four values of \(s_\mathrm{0,n}\). One sees that the two extreme values (\(s_\mathrm{0,n} = 6\) and \(9\,k_\mathrm{B}/\text{baryon}\)) are not representative of the predictions of the formation simulations, but are clearly too low/high in comparison (one should here keep in mind the logarithmic nature of the entropy. It means that a numerically small difference in \(s_\mathrm{0,n}\) actually corresponds to a very significant change of the gravothermal heat content). As visible from Fig.~\ref{fig:initialentro}, in the formation simulations there are, in particular, virtually no low- and intermediate-mass planets with \(s_0\) as low (high) as 6 (9) \(k_\mathrm{B}/\text{baryon}\).}  

\subsection{ {An example case}}
\begin{figure*}
	\centering
    \includegraphics[width=\linewidth]{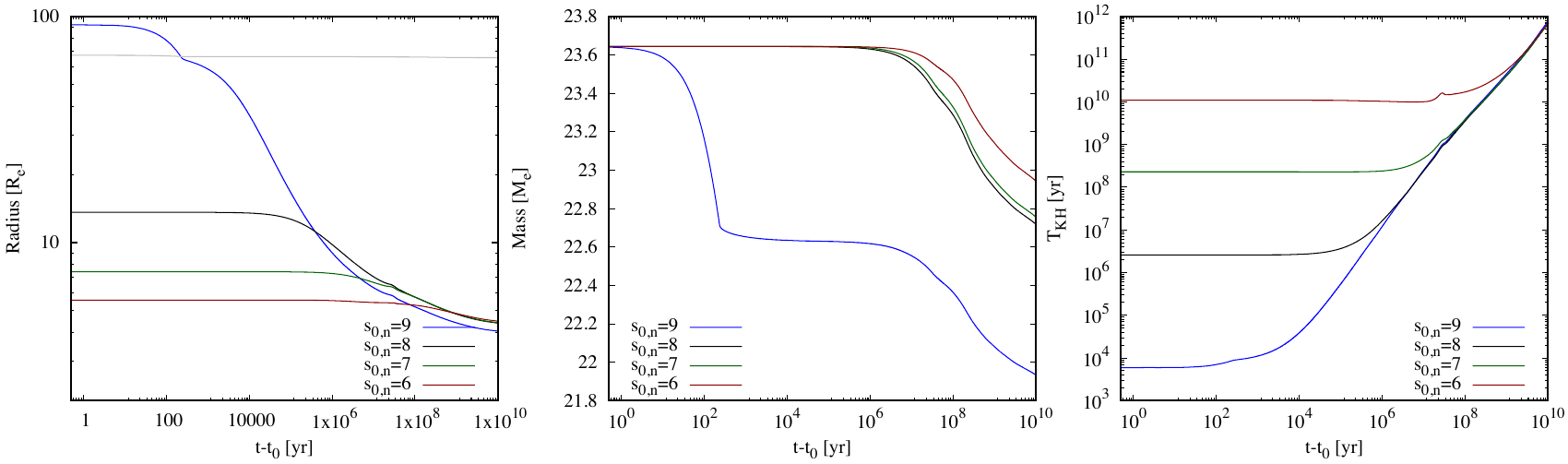}
    \caption{ {Temporal evolution of the outer radius (left), total mass (middle), and Kelvin-Helmholtz timescale (right panel) for a \(23.64\,M_\oplus \) planet (\(M_\text{core} = 20 M_\oplus \)) at \(0.1\,\mathrm{AU}\) for the four \(s_{0,n}\) indicated in the plots. Time is measured relative to the moment when the simulation starts. In the left panel, the grey line shows the Hill sphere radius. The planet with the (unrealistically) high \(s_{0,n} = 9\,k_\mathrm{B}/\text{baryon}\) initially overflows the Hill sphere, leading to a strong reduction of the envelope mass. At late times, this leads to a smaller radius. The other three cases which lack this overflow phase given in contrast similar values, with a slight anti-correlation of the radius at late time and the initial entropy.}   
    }\label{fig:case1000}
\end{figure*}
 {Figure~\ref{fig:case1000} shows the temporal evolution of a specific planet from the rectangular grid for the four \(s_\mathrm{0,n}\). The energy-/recombination-limited escape model is used, but qualitatively equivalent effects are also occurring for the hydrodynamic model. This planet has an orbital distance of \(0.1\,\mathrm{AU}\), \(M_\text{core} = 20 M_\oplus \), and \(M_\mathrm{env,0} = 3.64 M_\oplus \). The \(L_0\) is 0.34, 12.1, 602.4, and \(56104.3\,L_\mathrm{\jupiter}\) for the four \(s_0\) studied. The latter value is certainly extremely high for a planet of this mass \citep{Mordasini2017}. Equation~\ref{eq:l0par} yield for comparison \(14.3\,L_\mathrm{\jupiter}\).}

 {This planet was chosen because it illustrates with a specific example the two main findings of the grid analysis of the valley location as a function of \(s_\mathrm{0,n}\) in the next section: namely a weak impact of \(s_\mathrm{0,n}\) for the three lower entropy values, and envelope overflow for some planets for \(s_\mathrm{0,n} = 9 \, k_\mathrm{B}/\text{baryon}\).}

 {In the left panel we see the radius as a function of time. The initial radius is as expected the larger the higher \(s_\mathrm{0,n}\) is. This has the well-known consequence (e.g., \citealt{Owen2020}) that stronger XUV-driven atmospheric escape will occur for a higher \(s_0\) at young ages, such that at high ages, when the initial \(s\) if forgotten, the planet will have a smaller radius because it has a lower mass envelope. For the highest \(s_0\) case, there is, however, an additional effect: the huge initial radius is here larger than \(R_\text{roche}\), meaning that some envelope gas is unbound. In the model, we then remove at each timestep one third of the mass outside of \(R_\text{roche}\). This factor smaller than unity (which would in principle be the value to use) was chosen for numerical stability. The exact value is, however, inconsequential: in any case, extremely rapid and strong mass loss occurs until the outer radius becomes smaller than \(R_\text{roche}\), and only the time duration until this happens varies somewhat with the specific fraction chosen. On the other hand, in a situation of such rapid mass loss like here, quantitative results of our 1D strictly hydrostatic model with a radially constant luminosity at a given time should be regarded with caution.}

 {As is visible in the middle panel, this overflow phase removes about one third of \(M_\mathrm{env,0}\) on an extremely short timescale which is on the order of just 100 years. At late times, this Roche lobe overflow has the consequence that the planet has a clearly smaller radius (\(4.05\,R_\oplus \)) and mass than the other three cases. For them, the radius varies only between 4.39 and \(4.47\,R_\oplus \). The largest radius corresponds to the lowest \(s_0\) because this planet suffered from less ``normal'' XUV-driven escape because of its higher mean density at young ages, as mentioned.}

\begin{figure*}
	\centering
    \includegraphics[width=\linewidth]{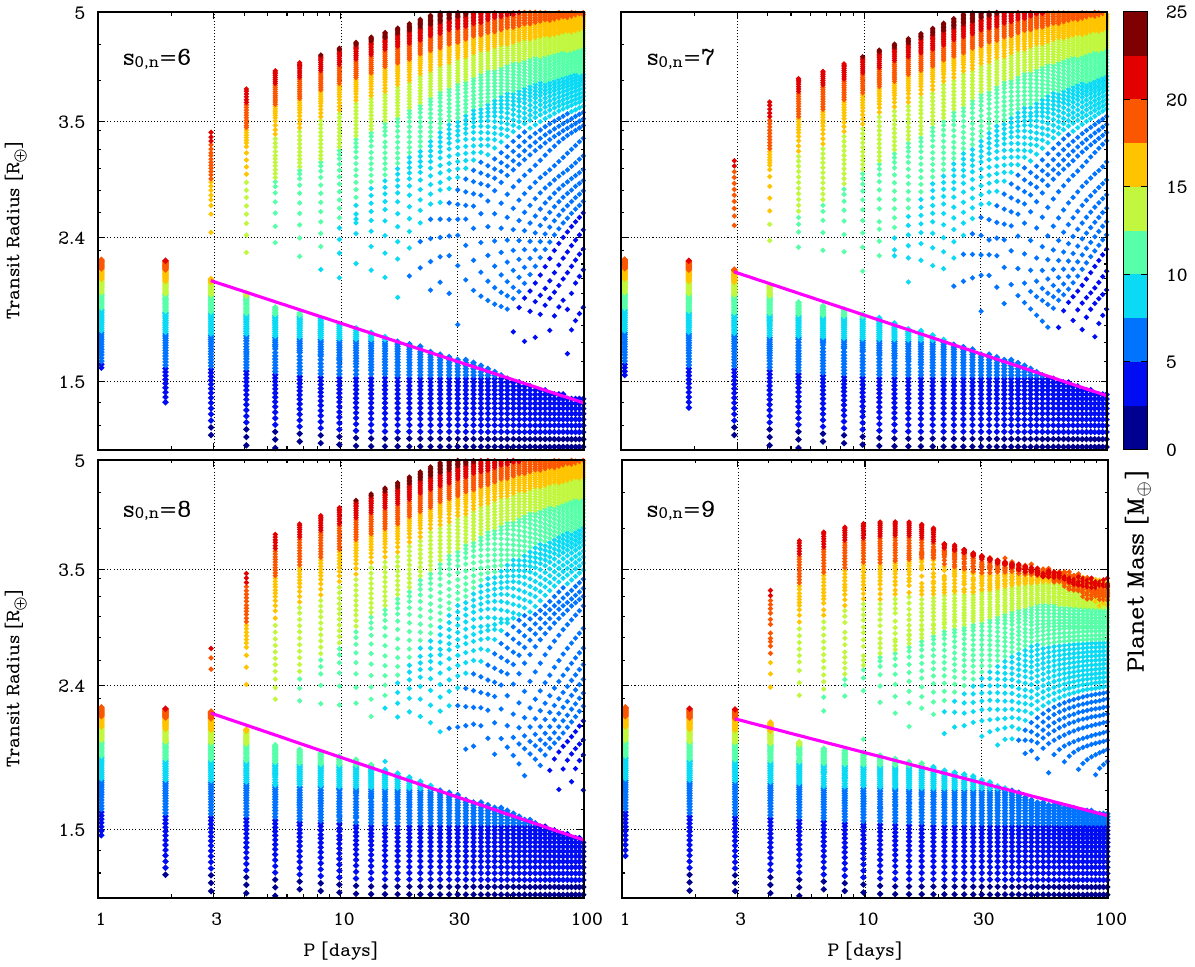}
    \caption{ {Transit radius as a function of orbital period at \(5\,\mathrm{Gyr}\) for the four initial entropies \(s_\mathrm{0,n} = 6\), \(7\), \(8\), and \(9\,k_\mathrm{B}/\text{baryon}\). The lower three values lead to virtually identical valley slopes. For the highest initial entropy, unstable initial conditions with Roche lobe overflow result, which strongly removes mass especially at the larger orbital distances. This affect the valley slope. Such a high entropy is, however, not in agreement with the predictions of formation models.}
    }\label{fig:gridfours0}
\end{figure*}

 {The occurrence of such an overflow phase is indicative of an unrealistic combination of initial conditions for the evolutionary phase in terms of core mass, envelope mass, and luminosity. In reality, during the precedent formation phase, while embedded in the nebula, a core of such a high luminosity (caused for example by a burst of solid accretion) would not posses an envelope of this mass. Instead, potential excess gas would get expelled out of the Hill sphere back into the surrounding disk, and \(M_\text{env}\) would be lower than assumed here. This effect is by construction not taken into account when \(s_0\) is assumed to be only a function of the total mass, as it is the case both for Eq.~\ref{eq:l0par} and~\ref{eq:s0par}. In the formation simulations solving the internal structure equations, this is in contrast automatically taken into account. Thus, whenever possible, \(s_0\) should be estimated in evolutionary models not only based on the total planet mass, but the core and envelope mass separately. Such a prescription for \(L_0(M_\text{core}, M_\text{env})\) can be found in \citetalias{2020A&A...638A..52M}.}

 {The right panel of Fig.~\ref{fig:case1000} shows the Kelvin-Helmholtz timescale. It is calculated with the actual numerically obtained total energy of the planets and not the approximation \(G M_p^2/R_p\). This approximation can yield very different incorrect values for planets with a very large, extremely tenuous outer envelope, as it is the case here at early times. We see that \(s_\mathrm{0,n} = 9 \, k_\mathrm{B}/\text{baryon}\) corresponds to an extremely small \(T_\mathrm{KH}\) of less than \(10^4 \, \mathrm{yr}\). A planet would thus extremely quickly evolve away from such conditions, making it an unlikely state to exist exactly at the moment of disk dissipation. The lowest \(s_\mathrm{0,n} = 6 \, k_\mathrm{B}/\text{baryon}\) on the other hand yields an extremely long initial \(T_\mathrm{KH} \mathbin{\sim} 10^{10} \, \mathrm{yr}\). The radius hardly changes for about 1 Gigayear. Such an extremely cold start seems also unlikely given the energy liberated when accreting a solid core of \(20 \, M_\oplus \).}

\subsection{ {Valley locus as function of \(s_0\)}}
 {The example of this individual planet suggests that the impact of \(s_0\) should be rather small, except if an unexpected high entropy is used. Figure~\ref{fig:gridfours0} showing the rectangular grid of simulations indeed reflects a similar pattern. The figure, which can be compared with Fig.~\ref{fig:grid} using the nominal \(L_0\), shows for the four \(s_{0,n}\) the radius at \(5\,\mathrm{Gyr}\) as a function of orbital period, colour-coding the total mass. The hydrodynamic escape model is used. As for the nominal rectangular grid (Sect.~\ref{sec:results}), we have made a least square fit to determine the valley slope \(\alpha \) and the normalisation radius at 10 days period, \(\tilde{R}_b\). The values for the hydrodynamic model are given in Table~\ref{tab:slopeentropy}. The result for the energy-/recombination-limited model are similar.}

\begin{table} 
    \centering
     {
    \caption{
        Valley bottom radius at an orbital period of 10 days \(\tilde{R}_\mathrm{b}\) and valley power law slope \(\alpha \) as a function of orbital period for the hydrodynamic evaporation model and the four initial entropies \(s_\mathrm{0,n}\).
    }\label{tab:slopeentropy}
    \begin{tabular}{lll}
        \hline\hline
        \(s_\mathrm{0,n}\) & \(\tilde{R}_\mathrm{b}\) [\(R_\oplus \)] & Slope \(\alpha \) \\
        \hline 
        6  
        & 1.81                  & \(-0.113\) \\
        7 
        & 1.86                  & \(-0.114\) \\
        8 
        & 1.90                  & \(-0.117\) \\
        9
        & 1.93                  & \(-0.089\) \\
        \hline
    \end{tabular}
    }
\end{table}

 {The panel in the bottom left corner of Fig.~\ref{fig:gridfours0} differs clearly from the other three, which are in contrast similar to each other. We see an absence of planets in the upper right corner. The iso-mass curves visible through the colour-code are significantly shifted, especially at larger orbital distances. The valley slope is also less steep than in other three cases. These differences are the consequence of intense mass loss resulting from Roche lobe overflow right at the beginning of the simulations, as seen in the example case. It affects both planets far from the valley (as in the example), but also planets close to it. {An interesting point is here that the maximum radii are limited to about 3.5 to 4 $R_{\oplus}$ which corresponds to the radius above which observationally the  frequency of planets drop strongly (the cliff, \citealt{Kite_2019}). This suggest that in the very high entropy case, it is not necessary to fine-tune the initial (i.e., post-formation) H/He masses to reproduce the cliff. This echoes the suggestion of \citet{2016ApJ...817..107O} that  the ``boil-off'' process could be partially responsible for the lack of larger planets.}
 
{Another small feature visible in Fig.~\ref{fig:gridfours0} is the absence of points in the bottom left corner. This is a consequence of the following: for these very close-in, low-mass planets, no structure was found for the requested \(s_0\). Because of Eq.~\ref{eq:menve0}, these planets have tenuous atmospheres approaching for lower \(s_0\) an isothermal structure. The given equilibrium temperature then excludes certain combinations of core mass, envelope mass, and \(s_0\). This is in contrast to colder and more massive planets with an inner convective zone.}

 {As discussed in the previous section, our quantitative results for planets undergoing Roche lobe overflow should be taken with caution, given our model's capabilities. However, this process only affects planets with \(s_\mathrm{0,n} = 9 \, k_\mathrm{B}/\text{baryon}\), which is not a likely initial condition for low-mass planets. The important conclusion from examining the role of \(s_0\) is rather the following: for the relevant, very wide range of \(s_\mathrm{0,n}\) from \(6\) to \(8 \, k_\mathrm{B}/\text{baryon}\), the impact of the post-formation entropy on the final valley slope is only very small, as can be seen in Table~\ref{tab:slopeentropy}. The slope \(\alpha \) hardly changes with values between \(-0.113\) and \(-0.117\). This is also the same as found for the nominal \(L_0\). We do see that \(\tilde{R}_b\) shifts to higher values with increasing \(s_0\) as expected, but the shift from \(1.81\) to \(1.90 \, R_\oplus \) is rather small. This is especially the case when one considers that formation models predict a spread of about only \(1\), and not \(3 \, k_\mathrm{B}/\text{baryon}\) at fixed planet mass.}

 {To summarise, we find that varying \(s_\mathrm{0,n}\) over a wide range of \(6\) to \(8 \, k_\mathrm{B}/\text{baryon}\) has virtually no impact on the valley slope, and shifts \(\tilde{R}_b\) only by a rather small increment. This range of initial entropies includes those suggested by formation models and leads to stable initial conditions where the initial planet radius is smaller than the Roche lobe. Only when using a for this mass range unrealistically high \(s_\mathrm{0,n} = 9\,k_\mathrm{B}/\text{baryon}\), the impact becomes significant, because mass loss via Roche lobe overflow occurs immediately at the beginning of the simulations. Such unstable initial conditions are, however, not predicted by formation models.}

\end{document}